\documentclass[journal=jpclcd,manuscript=article]{achemso}   	

\usepackage[]{standalone}
\usepackage{import}

\usepackage{xcolor}
\usepackage{float}

\usepackage{2IDP_HSC}

\title{\maintitle}

\author{Alan N. Amin}
\affiliation
{Department of Biochemistry, University of Toronto, Toronto, Ontario, Canada}
\altaffiliation{Contributed equally to this work}
\altaffiliation{Present address: Systems, Synthetic, and Quantitative Biology Program,\\ Harvard Medical School, Boston, Massachusetts, U.S.A.}
\author{$\null\hskip -1.8mm ^{,\S}$\ Yi-Hsuan Lin}
\affiliation
{Department of Biochemistry, University of Toronto, Toronto, Ontario, Canada}
\alsoaffiliation
{Molecular Medicine, Hospital for Sick Children, Toronto, Ontario, Canada}
\altaffiliation{Contributed equally to this work}
\author{Suman Das}
\affiliation
{Department of Biochemistry, University of Toronto, Toronto, Ontario, Canada}
\author{Hue Sun Chan}
\affiliation
{Department of Biochemistry, University of Toronto, Toronto, Ontario, Canada}
\alsoaffiliation
{Department of Molecular Genetics, University of Toronto, Toronto, Ontario, Canada}
\email{huesun.chan@utoronto.ca}

\date{\today}	

\begin{document}

\begin{abstract}

\noindent
Intrinsically disordered proteins (IDPs) are important for biological
functions. In contrast to folded proteins, molecular recognition among 
certain IDPs is ``fuzzy'' in that their binding and/or phase separation 
are stochastically governed by the interacting IDPs' amino acid sequences 
while their assembled conformations remain largely disordered.
To help elucidate a basic aspect of this fascinating yet poorly understood 
phenomenon, the binding of a homo- or hetero-dimeric pair of 
polyampholytic IDPs is modeled statistical mechanically using 
cluster expansion.  We find that the binding affinities of binary fuzzy 
complexes in the model correlate strongly with a newly derived
simple ``jSCD'' parameter readily calculable from the pair of IDPs' 
sequence charge patterns. Predictions by our analytical theory are 
in essential agreement with coarse-grained explicit-chain simulations.
This computationally efficient theoretical framework 
is expected to be broadly applicable to rationalizing and predicting 
sequence-specific IDP-IDP polyelectrostatic interactions.

\begin{tocentry}
\includegraphics[width=1.09\columnwidth]{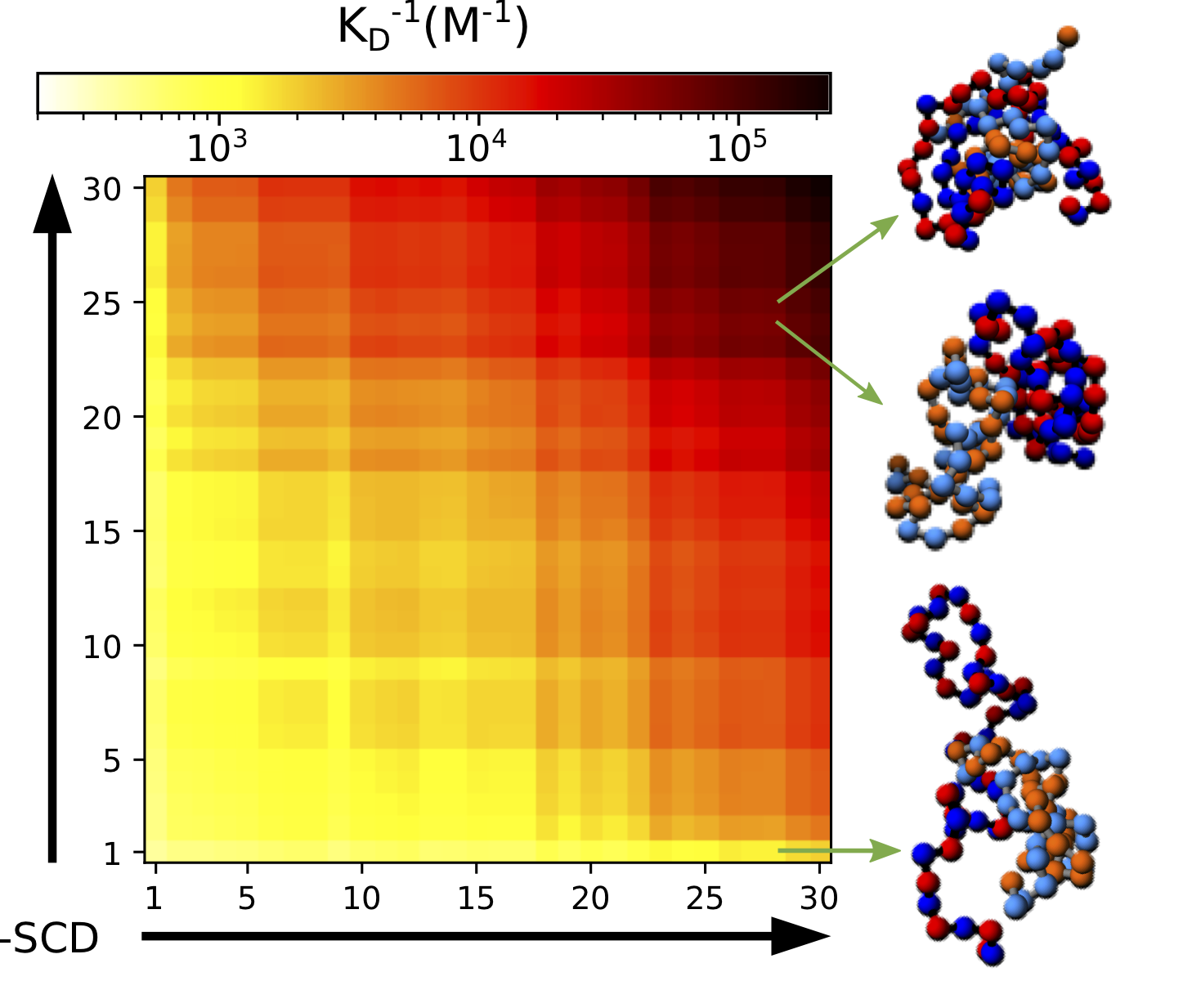}
\end{tocentry}

\end{abstract}

\maketitle

\section{Introduction}

Intrinsically disordered proteins (IDPs)---hallmarked by their lack of folding 
to an essentially unique conformation in isolation---serve many 
physiological functions.\cite{dunker_etal2001,vanderlee_etal2014} Compared 
to globular proteins, IDPs are depleted in nonpolar but enriched in polar, 
aromatic, and charged residues.~\cite{uversky2002} Some IDPs adopt 
ordered/folded conformations 
upon binding to folded targets~\cite{Wright2009} or 
after posttranslational modifications\cite{Bah2014}, others remain disordered. 
Among the spectrum of diverse possible behaviors\cite{julie2012}, the IDPs 
in certain IDP-folded protein complexes can be highly disordered, 
as typified by the kinase inhibitor/ubiquitin ligase Sic1-Cdc4
complexes \cite{borg07,Mittag2008}.

Complexes with bound IDPs that are disordered are aptly named ``fuzzy 
complexes''\cite{tompa2008,Fuxreiter2015,Fuxreiter2017,Pons2018}. 
The role of these IDPs' amino acid sequences 
in molecular recognition varies, depending on the situation. For Sic1-Cdc4, 
most of the charges in the disordered Sic1 probably take part in modulating
binding affinity via multiple spatially long-range electrostatic---termed 
polyelectrostatic\cite{borg07}---interactions with the folded Cdc4 without 
locally engaging the Cdc4 binding pockets~\cite{Mittag2008,veronika17}. In 
contrast, for the IDP transactivation domain of Ewing sarcoma, 
sequence-dependent oncogenic effects may be underpinned largely by 
multivalent, spatially short-range polycation-$\pi$ interactions implicating
the IDP's tyrosine residues~\cite{KAW,cosb15}. More broadly, for 
multiple-component phase separation of IDPs, a ``fuzzy'' mode of molecular 
recognition was proposed whereby mixing/demixing of phase-separated 
polyampholyte species depends on quantifiable differences in the IDPs' 
sequence charge patterns~\cite{Lin2017c}. Variations aside, 
these mechanisms share the commonality of being stochastic in
essence, involving highly dynamic conformations, and as such are distinct 
from those underlying the structurally specific and relatively static 
binding participated by folded proteins. We thus extend the 
usage of ``fuzzy'' as an adjective not only for the structural features of 
certain biomolecular assemblies but also for the molecular recognition 
mechanisms that contribute to the formation of fuzzy assemblies.
This concept is applicable to nonbiological polymers as well.
Whereas multivalency, stochasticity, and conformational diversity have
long been the mainstay of polymer physics, recently,
sequence specificity and therefore fuzzy molecular recognition
has become increasingly important for nonbiological 
heteropolymers because of experimental advances in ``monomer precision'' 
that allows for the
synthesis of sequence-monodisperse polymers~\cite{PerrySing2020}.

As far as biomolecules are concerned, fuzzy molecular recognition should 
play a dominant
role in ``binding without folding'' IDP complexes wherein the bound
IDPs are disordered~\cite{sigalov2016,schuler_cosb_2020}. 
Generally speaking, a condensed
liquid droplet of IDP is a mesoscopic fuzzy assembly underpinned 
by a fuzzy molecular recognition mechanism.\cite{Lin2017c} 
With regard to basic binary (two-chain) IDP complexes, evidence has 
long pointed to their existence,\cite{sigalov_etal2007,kragelund2008}
although extra caution needed to be used to interpret the pertinent 
experimental data.\cite{tanja2014,sigalov2016}
Of notable recent interest is the interaction between
the strongly but oppositely charged H1 and ProT$\alpha$ IDPs
involved in chromatin condensation and remodeling, which
remain disorder while forming a heterodimer 
with reported dissociation constant ranging from 
nanomolar~\cite{Borgia2018} to sub-micromolar levels~\cite{Feng2018}.


We now tackle a fundamental aspect of fuzzy molecular recognition, namely the 
impact of sequence-specific electrostatics on binary fuzzy 
complexes. Electrostatics is important for IDP 
interactions\cite{borg07,Borgia2018,YQHuang2020,schuler_cosb_2020} 
including phase 
separation.\cite{Lin2017c, Wang2018, Tsang2019} IDP sequence specificity is 
a key feature of their single-chain properties\cite{Das2013,Sawle2015,Lin2017b,Zheng2020} 
and multiple-chain phase 
behaviors.\cite{Nott2015,Lin2016,Pak2016,Lin2017a,Lytle2017,SingPerry2017,Das2018b,McCarty2019,panag2020,jeetain2020} 
IDP properties depend not only on their net charge but are also
sensitive, to various degrees, to their 
specific sequence charge pattern, which has been characterized by two
parameters, $\kappa$ and ``sequence charge decoration'' (SCD):
$\kappa$ is an intuitive blockiness measure\cite{Das2013}; whereas
\begin{equation}
\SCD(\{ \ch{} \}) = \frac{1}{2N} \sum_{s,t =1}^{N}\ch{}_s\ch{}_t\sqrt{|s-t|}
\; ,
\end{equation}
defined for any charge sequence 
$\{ \ch{} \} = \{ \ch{}_1, \ch{}_2, \dots,\ch{}_N \}$,
emerges from an analytical theory for polyampholyte dimensions\cite{Sawle2015}.
Both single-chain dimensions\cite{Das2013,Sawle2015,Lin2017b} and phase 
separation 
propensities\cite{Lin2017b,Das2018a,Das2018b,McCarty2019,jeetain2020} are
seen to correlate with these parameters. These measures are found 
to be evolutionarily conserved among IDPs, suggesting intriguingly that the 
gestalt properties they capture are functionally significant~\cite{Zarin2019}.
Our present focus is on binary complexes, which are of interest 
themselves\cite{Borgia2018} and possibly also as proxies for mesoscopic 
multiple-IDP phase behaviors. Generalizing such a correspondence
between single-chain properties and multiple-chain phase
behaviors for homopolymers~\cite{Panagio1998,WangZG2014} and
polyampholytes~\cite{Lin2017b}, for example,
the osmotic second virial coefficient, 
$\btwo$, of a pair of IDP chains has been proposed as an approximate 
measure for the IDP's sequence-dependent 
phase separation propensity~\cite{Dignon2018}.


\section{Methods}

With these thoughts in mind, we develop an analytical theory for binary 
IDP-IDP electrostatics. As exemplified by recent 
studies of phase behaviors,\cite{Lin2016,Lin2017a,Kings} approximate analytical 
theories, among complementary approaches, are conceptually productive
and efficient for gaining insights into 
sequence-specific IDP behaviors. The system analyzed herein consists of two 
IDPs $A$, $B$ of lengths $\NA$, $\NB$; charge sequences 
$\{ \ch{A} \} = \{ \ch{A}_1, \ch{A}_2,  \dots,\ch{A}_{\NA} \}$,
$\{ \ch{B} \} = \{ \ch{B}_1, \ch{B}_2, \dots, \ch{B}_{\NB} \}$;
and residue (monomer) coordinates 
$\{ \R^A \} = \{ \R_1^A, \R_2^A,\dots, \R_{\NA}^A \}$, 
$\{ \R^B \} = \{ \R_1^B, \R_2^B,\dots, \R_{\NB}^B \}$. 
Both $A=B$ (homotypic) and $A\ne B$ (heterotypic) cases are considered.
Key steps in the formal development are presented below;
details are provided in the Supporting Information.
The second virial coefficient of the IDP pair is given by~\cite{Pathria}
\begin{equation}
\btwo = \int d\Rcm^{AB}\; 
\Avg{ 1- e^{-\beta \U^{AB}(\Rcm^{AB}; \{ \R^A\}, \{\R^B\})}  }_{A,B}
\; ,
	\label{eq:btwo_Rcm}
\end{equation}
where $\beta=1/\kB T$ ($\kB$ is Boltzmann constant, $T$ is absolute
temperature), $\Rcm^{AB}$ is the center-of-mass distance and
$\U^{AB}$ is the total interaction between $A$ and $B$, and
the average $\Avg{\cdots}_{A,B}$ is over the conformational ensembles 
of $A$ and $B$. To simplify notation, we use $\U^{AB}$ to denote
$\beta \U^{AB}$ below. Now,
Eq.~\ref{eq:btwo_Rcm} may be rewritten as
\begin{equation}
\btwo 
= \Vol-\frac{\QAB}{\QA\QB}
= \Vol \int \DD[\R^A] \DD[\R^B]\PP{A}[\R^A]\PP{B}[\R^B]\left( 
1- e^{-\U^{AB}[\R^{A}, \R^B]}  \right)
\; ,
	\label{eq:B2_integral}
\end{equation}
where $\Vol$ is volume, 
$\QAB$ is the partition function of the entire $A$-$B$ system;
$\QA$ and $\QB$ are, respectively, the isolated single-chain 
partition functions of $A$ and $B$,
$\DD[\R^i] \equiv \int \prod_{s=1}^{N_i}d\R^i_s$ with $i = A, B$,
and $\PP{i}[\R^i]$ is 
the single-chain probability density for conformation $\{ \R^i \}$.
Note that in the limiting case with no internal degrees of
freedom in $A$ and $B$, i.e., when $\NA=\NB=1$, both Eqs.~\ref{eq:btwo_Rcm} 
and \ref{eq:B2_integral} 
reduce to $\btwo = \int d^3 r \{1-\exp[-\U_{AB}(r)]\}$.

%

When $\U^{AB}$ is a sum 
of pairwise interactions between residues in different polymers:
\begin{equation}
\U^{AB}[\R^A, \R^B] = 
\sum_{s=1}^{\NA}\sum_{t=1}^{\NB}\Vr^{AB}_{st}\left(\R^{AB}_{st}\right)
\; ,
\end{equation}
where $\R^{AB}_{st} \equiv \R^{A}_s - \R^B_t$ and 
$\Vr^{AB}_{st}$ is the $(s^A,t^B)$ potential energy between
the $s$-th residue in $A$ and the $t$-th residue in $B$,
the integrand in Eq.~\ref{eq:B2_integral} may be expressed
as a cluster expansion: 
\begin{equation}
\begin{aligned}
e^{-\U^{AB}[\R^A, \R^B]} -1
= & \left\{ \prod_{s=1}^\NA \prod_{t=1}^\NB 
	  \left[ \left( e^{-\Vr^{AB}_{st}(\R^{AB}_{st})}-1\right)+1\right] \right\} 
- 1 \\
= & \sum_{s=1}^{\NA} \sum_{t=1}^{\NB} f_{st} 
+ \sum_{s\geq t=1}^{\NA} \sum_{l\geq m=1}^{\NB} f_{sl} f_{tm} 
- \sum_{s=1}^{\NA}\sum_{t=1}^{\NB}f_{st}^2
+ O\left(f^3\right)
\; ,
    \label{eq:expUAB-f}
\end{aligned}
\end{equation}
where $f_{st} \equiv \exp[-\Vr^{AB}_{st}(\R^{AB}_{st})]-1$ is the Mayer 
$f$-function for $(s^A,t^B)$.  
Intuitively, the first and third terms of the last expression in 
Eq.~\ref{eq:expUAB-f} are functions of $f_{st}$ which involves only one
residue per chain and thus is independent of the $\PP{i}$s for
relative positions along the same chain. In contrast, the second term of 
$f_{st}f_{lm}$ involves two pairwise interchain interactions and thus 
$\PP{i}$-governed correlation of same-chain residue positions.
Defining the Fourier transformed ($\kk$-space) matrices 
of intrachain residue-residue correlation function 
\begin{equation}
\left[\hat{P}^i(\kk)\right]_{st} \equiv \int \DD[\R^i] \PP{i}[\R^i] e^{i\kk\cdot\left(\R^i_s-\R^i_t\right)} \; , \; i = A, B,
	\label{eq:Pk}
\end{equation}
and of the Mayer $f$-function
\begin{equation}
\left[\hat{f}(\kk)\right]_{st} \equiv \int d\rr f_{st}(\rr)e^{i\kk\cdot\rr}
\; ,
	\label{eq:fst_k}
\end{equation}
the $O(f^2)$ cluster expansion of $\btwo$ is derived in
the Supporting Information as
\begin{equation}
\btwo =
    - \sum_{s=1}^{\NA}\sum_{t=1}^\NB\left[\hat{f}(\zero)\right]_{st} 
    - \frac{1}{2}\int \frac{d^3k}{(2\pi)^3} 
        {\rm Tr}\left[ 
            \hat{f}(\kk)\hat{P}^B(-\kk)\hat{f}^{\rm T}(-\kk)\hat{P}^A(-\kk) 
            - \hat{f}(\kk)\hat{f}^{\rm T}(-\kk) \right] + O(f^3)
\; ,
    \label{eq:B2_general}
\end{equation}
where the ``${\rm T}$" superscript of a matrix denotes its transpose.
Focusing on electrostatics, we first consider a screened Coulomb potential,
$\Vr^{ij}_{st}(r) = \lb \ch{i}_s\ch{j}_t \exp(-\kappaD r)/r$, which
is equivalent to
\begin{equation}
\left[\hat{\Vr}(\kk)\right]_{st} =
\frac{4\pi\lb}{k^2+\kappaD^2}\ch{i}_s\ch{j}_t
\end{equation}
in $\kk$-space, where $\lb=e^2/(4\pi\epsilon_0\epsilon_{\rm r} \kB T)$ 
is Bjerrum length, $\epsilon_0$ and $\epsilon_{\rm r}$ are vacuum and
relative permittivity, respectively,
$\kappaD$ is Debye screening wave number (not to be confused with
the sequence charge pattern parameter $\kappa$). 
The case of pure Coulomb interaction (without screening) will be considered
below.
We then make two approximations in Eq.~\ref{eq:B2_general} for tractability. 
First, we approximate the IDP conformations as 
Gaussian chains with Kuhn length $b$ (Ref.~\citen{Lin2017a}),
\begin{equation}
\left[\hat{P}^i(\kk)\right]_{st} 
    \approx \left[ \GaussianPM{}{\kk}\right]_{st}= 
    e^{-\frac{1}{6}(kb)^2|s-t| }
\label{eq:Gaussian}
\end{equation}
where $k\equiv|\kk|$.
Second, we express the Mayer $f$-functions as high-temperature expansions:
\begin{equation}
\left[\hat{f}(\kk)\right]_{st}  
=  -\frac{4\pi\lb}{k^2+\kappaD^2}\ch{i}_s\ch{j}_t 
+ \frac{2\pi\lb^2}{k}\left(\ch{i}_s\ch{j}_t\right)^2\tan^{-1}\left(
\frac {k}{2\kappaD}\right) + O(\lb^3)
\; .
\end{equation}
With these two approximations, $\btwo$ up to $O(\lb^2)$ is given by
\begin{equation}
\btwo \approx
    \frac{4\pi\lb}{\kappaD^2}\nch{A} \nch{B}
    -4\lb^2 \int \frac{dk k^2}{(k^2+\kappaD^2)^2}
    \sum_{s,t=1}^\NA\sum_{l,m=1}^\NB
        \ch{A}_s\ch{A}_t\ch{B}_l\ch{B}_m
        e^{-\frac{1}{6}(kb)^2[|s-t|+|l-m|]}
\; ,
    \label{eq:B2_F1F2}
\end{equation}
where $\nch{i} \equiv \sum_{s=1}^{N_i}\ch{i}$ is the net charge of $i$.
The two terms account, respectively, for the mean-field Coulomb 
interaction between the two chains' net charges 
and sequence specificity. 
\\

\section{Results and Discussion}

{\bf\large Dominant role of disorder in salt-dependent IDP binding.} 
Let $\theta$ be the binding probability of chains $A,B$ with
the same concentration $[c]$. The probability that they 
are not bound 
\begin{equation}
1-\theta \equiv \frac{\Vol\QA\QB}{\QAB} = \frac{1}{1-\btwo/\Vol}
\;
\end{equation}
when $V$ is chosen, without loss of generality, to include only 
an $A,B$ pair and thus $[c]=1/V$ (cf. Eq.~\ref{eq:B2_integral}).
It follows that the dissociation constant $\KD$ is given by
\begin{equation}
\frac{1}{\KD} = \frac{\theta[c]}{(1-\theta)^2[c]^2} 
= -\btwo(1-\btwo/\Vol)
\approx  -\btwo
\; ,
    \label{eq:Kdeqn}
\end{equation}
where the last approximation holds at low $A,B$ concentrations.

To gain insight into the physical implications of the perturbative
terms in the $\btwo$ expression in
Eq.~\ref{eq:B2_F1F2}, we first apply them, through Eq.~\ref{eq:Kdeqn}, to the
binding of IDPs H1 and ProT$\alpha$ for which salt-dependent
$\KD$s have recently been measured experimentally~\cite{Borgia2018,Feng2018}. 
H1 and ProT$\alpha$ contain $\approx 110$ and $\approx 200$ residues, 
respectively, with small length variations for different constructs. 
We use the 202-residue H1 and 114-residue ProT$\alpha$ sequences 
in Table~\ref{tb:Kd_H1_ProTa} of the Supporting Information
for theoretical calculations, assigning $-1$ charge to each D and E 
residue, $+1$ charge to each R and K residue, and zero charge to other 
residues. We set $T=293.15$~K, which is equal\cite{Borgia2018,Feng2018} 
or similar\cite{Borgia2018} 
to those used for various experiments. As a first approximation, we
apply the standard 
relation $\kappaD=(8\pi\lb {\cal N}_{\rm A} [{\rm NaCl}])^{1/2}$,
where ${\cal N}_{\rm A}$ is Avogadro number and
$\epsilon_{\rm r}=78$ for bulk water, to model dependence
on NaCl concentration. It should be noted, however, that recent
experiment showed that an ``anomalous'' decrease in $\kappaD$
with increasing NaCl concentration likely ensues for [NaCl] $\gtrsim 500$ mM.
(Ref.~\citen{Perkin2016}).

\begin{figure}[h]
    \centering
    \includegraphics[width=0.5\columnwidth]{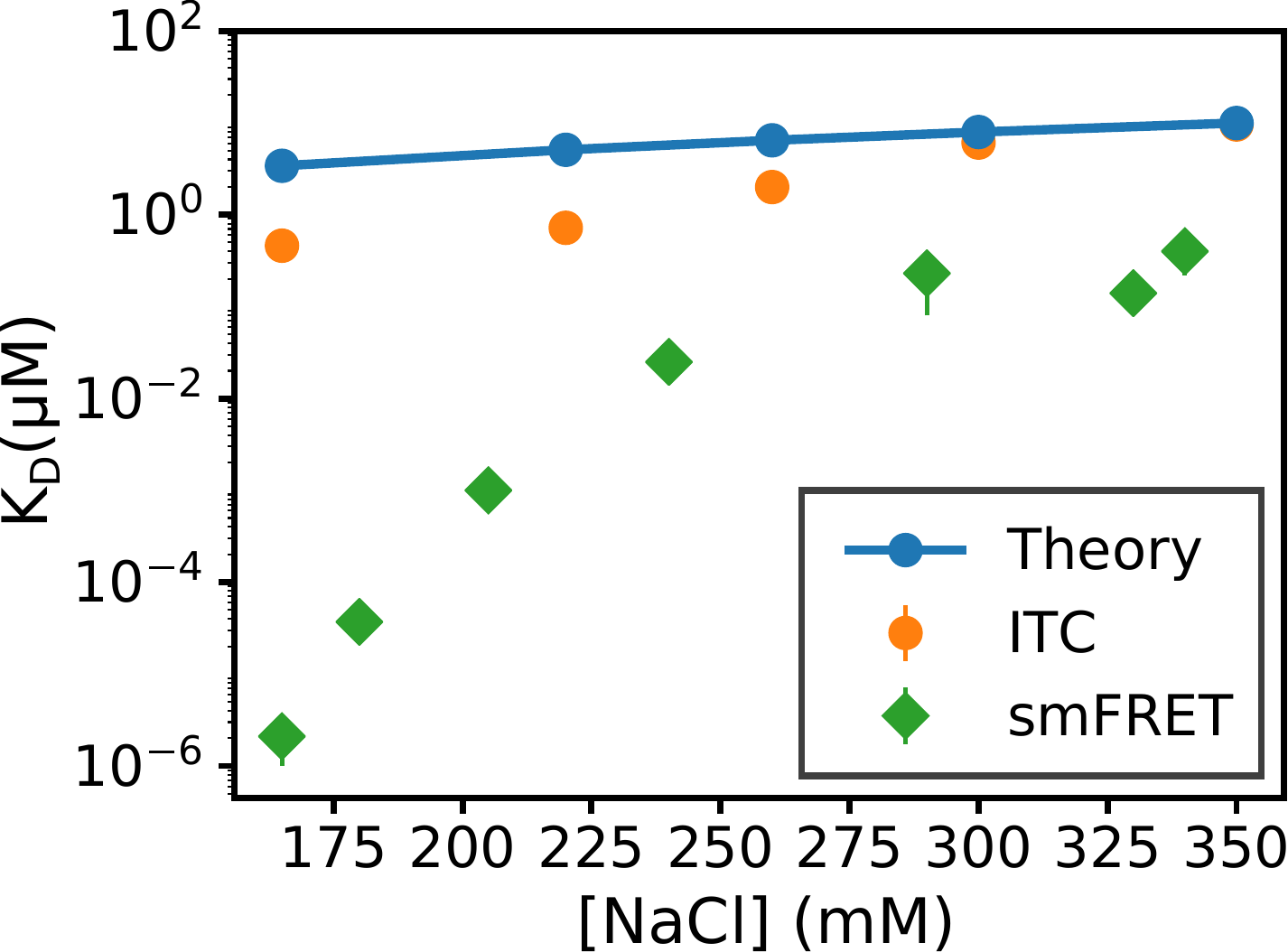}
    \caption{Theoretical and experimental H1-ProT$\alpha$ dissociation 
constants as functions of salt concentration. Data plotted are provided 
in Table~\ref{tb:Kd_H1_ProTa} in the Supporting Information.}
    \label{fig:Kd_H1_ProTa}
\end{figure}

The theoretical salt-dependent $\KD$s of H1 and ProT$\alpha$ thus calculated 
using Eqs.~\ref{eq:B2_F1F2} and \ref{eq:Kdeqn} are shown in
Fig.~\ref{fig:Kd_H1_ProTa} toegther with single-molecule F\"orster resonance 
energy transfer (smFRET)\cite{Borgia2018} and isothermal 
calorimetry (ITC)~\cite{Feng2018} experimental data.
All three set of data show decrease in $\KD$ (increase in binding) with 
decreasing salt, but there is a large difference between the smFRET 
and ITC data. Notably, when [NaCl] is decreased from $\approx 350$ to
$160$ mM, smFRET measured an $\approx 2\times 10^5$ whereas
ITC measured only an $\approx 20$ times increase in binding affinity. 
This discrepancy remains to be resolved, as a careful examination of
the experimental conditions is necessary, including the possible presence of 
not only binary H1-ProT$\alpha$ complexes but also oligomers in the sample 
used in the experiments~\cite{schuler_cosb_2020,schuler_abstract_2020}.

Our theoretical $\KD$s are within an order of magnitude of those measured
by ITC. They are practically identical at 350 mM [NaCl], but our
theoretical $\KD$ decreases only $\approx 3$ times at [NaCl] = 165 mM rather
than the $\approx 20$ times for ITC~\cite{Feng2018}. Our theory also
predicts weaker ProT$\alpha$ binding for the H1 C-terminal region than 
for full-length H1 (Fig.~\ref{tb:Kd_H1_ProTa}) as seen in smFRET experiment, 
but our predicted $\sim 1.5$ times increase in $\KD$ is less than 
the $\approx 20$ times measured by smFRET experiment~\cite{Borgia2018}.
In general, our cluster expansion (Eq.~\ref{eq:expUAB-f}),
which is a high-$T$ expansion~\cite{Pathria}, is
less accurate when electrostatic interaction is strong, 
such as at zero or low salt, because $\btwo$ in Eq.~\ref{eq:B2_F1F2}
includes only two terms in a perturbation series, neglecting attractive 
terms of order $\lb^3$ and higher.  This consideration 
offers a perspective to understand the modest difference between our 
theory and ITC measurement at low salt. However, although the partial agreement 
between theory and ITC is tantalizing, our current theory should be most
useful for conceptual and semi-quantitative investigation of comparative
sequence dependence of different IDP complexes rather than as a quantitative 
predictor for the absolute binding affinity of a particular pair of IDPs.
Our theory ignores many structural and energetic details for tractability, 
including ion condensation, the effect of which has a 
salt dependence~\cite{Muthu2017,Lytle2017} that might underlie the dramatic 
salt dependence of $\KD$ as seen by smFRET~\cite{Borgia2018}, and other 
solvation effects that might necessitate an effective separation-dependent 
dielectric~\cite{karplus1979,freed2008}. After all, explicit-chain simulation
has produced a $\KD\approx 7\times 10^{-9}$ $\mu$M which is $>300$ times
more favorable than that measured by smFRET~\cite{Borgia2018}, 
underscoring that, as it stands, all reported H1-ProT$\alpha$ experimental 
data are within theoretical possibilities.

Limitations of our analytical formulation notwithstanding,
an important physical insight is gained by inspecting the 
contributions in Eq.~\ref{eq:B2_F1F2} to the predicted 
H1-ProT$\alpha$ behavior from the first mean-field term 
that depends solely on overall net charges of the two IDPs and 
the second, sequence-specific term. Remarkably, the mean-field net-charge
term alone yields $\KD$s that are 30--40 times larger than those
calculated using both terms in Eq.~\ref{eq:B2_F1F2}
(Table~\ref{tb:Kd_H1_ProTa}), indicating that the net-charge term is
almost inconsequential and that the sequence-dependent
term---and by extension also the $O(\lb^3)$ terms---embodying the dynamic 
disorder of IDP conformations play a dominant role in the favorable assembly 
of fuzzy IDP complexes.


{\bf\large Assembly of binary fuzzy complex is highly sequence specific.}
We now proceed to compare the binding of different IDP pairs and analyze 
them systematically by expressing
the $\btwo$ for electrostatic interactions
in Eq.~\ref{eq:B2_F1F2} as 
\begin{equation}
\btwo \approx
    \frac{4\pi\lb}{\kappaD^2}\nch{A} \nch{B}
    -4\lb^2 \int \frac{dk k^2}{(k^2+\kappaD^2)^2}
    \sum_{s,t=1}^\NA\sum_{l,m=1}^\NB
        \ch{A}_s\ch{A}_t\ch{B}_l\ch{B}_m
        e^{-\frac{1}{6}(kb)^2[|s-t|+|l-m|]}
\equiv  F_1 + F_2
\; ,
    \label{eq:B2_F1F2_SI}
\end{equation}
where $F_1$ is an $O(\lb)$ term arising from the interaction between
the net charges $\nch{A}$ of chain $A$ and $\nch{B}$ of chain $B$,
and $F_2$ accounts for sequence specificity. We further rewrite $F_2$ as
\begin{equation}
\begin{aligned}
F_2 = & -4\lb^2\int_0^\infty \frac{dk k^2}{(k^2+\kappaD^2)^2}
    \sum_{s,t=1}^\NA\sum_{l,m=1}^\NB
    \ch{A}_s\ch{A}_t\ch{B}_l\ch{B}_m
    e^{-\frac{1}{6}(kb)^2(|s-t| + |l-m|)} \\
\equiv & -\frac{4\lb^2 b}{\sqrt{6}}
    \sum_{s,t=1}^\NA\sum_{l,m=1}^\NB
    \ch{A}_s\ch{A}_t\ch{B}_l\ch{B}_m
    I_{(|s-t|+|l-m|)}
\; ,
\end{aligned}
    \label{eq:F_2_ori}
\end{equation}
where $I$ is the following integral over the variable $\kb$:
\begin{equation}
I_X = \int_0^\infty \frac{d\kb\kb^2}{(\kb^2+\kpb^2)^2}
e^{-X\kb^2}
\label{eq:IX0}
\end{equation}
with $\kb \equiv kb/\sqrt{6}$, $\kpb \equiv \kappaD b/\sqrt{6}$,
and $X \equiv |s-t|+|l-m|$. Using integration by parts,
\begin{equation}
\begin{aligned}
I_X =
-\frac{1}{2}\int_0^\infty d\kb
(\kb e^{-X\kb^2}) \frac {d}{d\kb} \frac {1}{\kb^2+\kpb^2}
= & -\frac{1}{2}\left.\frac{\kb e^{-X\kb^2}}{\kb^2+\kpb^2}\right|_0^\infty
        + \frac{1}{2}\int_0^\infty d\kb \frac{1-2X\kb^2}{\kb^2+\kpb^2} e^{-X\kb^2} \\
= & \left(\frac{1}{2}+X\kpb^2\right)
    \int_0^\infty d\kb\frac{e^{-X\kb^2}}{\kb^2+\kpb^2}
    - X \int_0^\infty d\kb e^{-X\kb^2} \\
= & \left(\frac{\pi}{4\kpb}+\frac{\pi X\kpb}{2}\right)
    e^{X\kpb^2}{\rm erfc}\left(\kpb\sqrt{X} \right)
    - \frac{\sqrt{\pi X}}{2}
\; ,
\label{eq:IX}
\end{aligned}
\end{equation}
where ${\rm erfc}(z)=(2/\sqrt{\pi})\int_z^\infty dt \exp(-t^2)$ is
the complementary error function.\cite{Ng_Geller-1969}
In a $\kpb \ll 1$ regime that allows for the substitution of ${\rm erfc}(z)$
and $e^{z^2}$ by their Taylor series,
\begin{equation}
e^{z^2}{\rm erfc}(z) = 1 - \frac{2z}{\sqrt{\pi}} + z^2  + O(z^3)
\label{eq:erfc}
\; ,
\end{equation}
setting $z=\sqrt{X}\kpb$ and applying Eq.~\ref{eq:erfc} to the last
expression in Eq.~\ref{eq:IX} yields
\begin{equation}
I_X = \frac{\pi}{4\kpb} - \sqrt{\pi X} +  \frac{3}{4}\pi\kpb X + O(\kpb^2)
\; .
\end{equation}
In that case $F_2$ in Eq.~\ref{eq:B2_F1F2_SI} becomes
\begin{equation}
F_2 =  -\frac{4\lb^2 b}{\sqrt{6}}\left[
\frac{\pi\sqrt{6}}{4\kappaD b} \left(\nch{A}\right)^2\left(\nch{B}\right)^2
    - \sqrt{\pi}\sum_{s,t=1}^\NA\sum_{l,m=1}^\NB\ch{A}_s\ch{A}_t\ch{B}_l\ch{B}_m
        \sqrt{|s-t|+|l-m|}
\right] + O(\kpb)
\; ,
    \label{eq:F2_Taylor}
\end{equation}
where the first term is an $O(\lb^2)$ contribution due to the chains' net
charges $\nch{A}$ and $\nch{B}$, the second term involving individual
$\sigma^i_s$s then provides the lowest-order (in $\kpb$) account of
sequence specificity.
A two-chain sequence charge pattern parameter, which we refer to
as ``joint sequence charge decoration" (jSCD) because of its formal
similarity with the single-chain SCD (Ref.~\citen{Sawle2015}),
emerges naturally from this sequence-specific term in
Eq.~\ref{eq:F2_Taylor}:
\begin{equation}
\jSCD(\ch{A}, \ch{B}) \equiv -\frac{1}{2\NA\NB}
    \sum_{s,t=1}^\NA\sum_{l,m=1}^\NB
        \ch{A}_s\ch{A}_t\ch{B}_l\ch{B}_m
        \sqrt{|s-t|+|l-m|}.
    \label{eq:jSCD}
\end{equation}
When one or both of the chains are overall neutral, i.e.,
$\nch{A}=0$ and/or $\nch{B}=0$ ($\nch{A}\nch{B}=0$),
both $F_1$ and the first term of $F_2$ in Eq.~\ref{eq:F2_Taylor}
vanish, leaving $\btwo$ in a form that is is proportional to $\jSCD$:
\begin{equation}
\left.\btwo\right|_{\kappaD\to0, \nch{A}\nch{B}=0} =
-8\sqrt{\frac{\pi}{6}}\lb^2 b \NA\NB\times \jSCD(\ch{A},\ch{B})
\; .
    \label{eq:B2jSCD}
\end{equation}
When both chains are not overall neutral, i.e.,
$\nch{A}\neq0$ and $\nch{B}\neq0$ ($\nch{A}\nch{B}\neq 0$), the
$\nch{A}\nch{B}$ terms in Eqs.~\ref{eq:B2_F1F2_SI} and \ref{eq:F2_Taylor} are
part of the Taylor series of the Mayer $f$-function of the mean-field (MF)
net charge interaction, as can be seen from the identity of these terms
with the first two terms in the Taylor expansion of the second virial
coefficient (denoted $\btwo^{\rm MF}$ here) of two point charges
interacting via a screened Coulomb potential:
\begin{equation}
\begin{aligned}
\btwo^{\rm MF} = & \int d^3 r \left(1- e^{-\lb\nch{A}\nch{B}e^{-\kappaD r}/r} \right) \\
= & 4\pi \int_0^\infty dr r^2 \left(
    \lb\frac{\nch{A}\nch{B}e^{-\kappaD r}}{r}
    - \frac{\lb^2}{2}\frac{\left(\nch{A}\right)^2\left(\nch{B}\right)^2e^{-2\kappaD r}}{r^2}
    + O(\lb^3)
    \right) \\
= & \frac{4\pi\lb\nch{A}\nch{B}}{\kappaD^2}
    - \frac{\pi\lb^2}{\kappaD}\left(\nch{A}\right)^2\left(\nch{B}\right)^2 + O(\lb^3)
\; .
\end{aligned}
        \label{eq:b2_MF_show}
\end{equation}
Since these $\nch{A}\nch{B}$ terms in Eqs.~\ref{eq:B2_F1F2_SI} and
\ref{eq:F2_Taylor} do not involve individual $\sigma^i_s$s
and thus include no sequence specificity, the $\jSCD$ term is always
the lowest-order term (in $\kpb$) that takes into account sequence
specificity for overall neutral as well as overall non-neutral chains.
We also note that the divergence of these net charge terms
in the $\kappaD \to 0$ limit is the well-recognized infrared divergence
caused by the long-range nature of pure Coulomb interaction, which
is regularized as long as there is nonzero screening ($\kappaD > 0$).

\begin{figure}
    \centering
    \includegraphics[width=0.8\columnwidth]{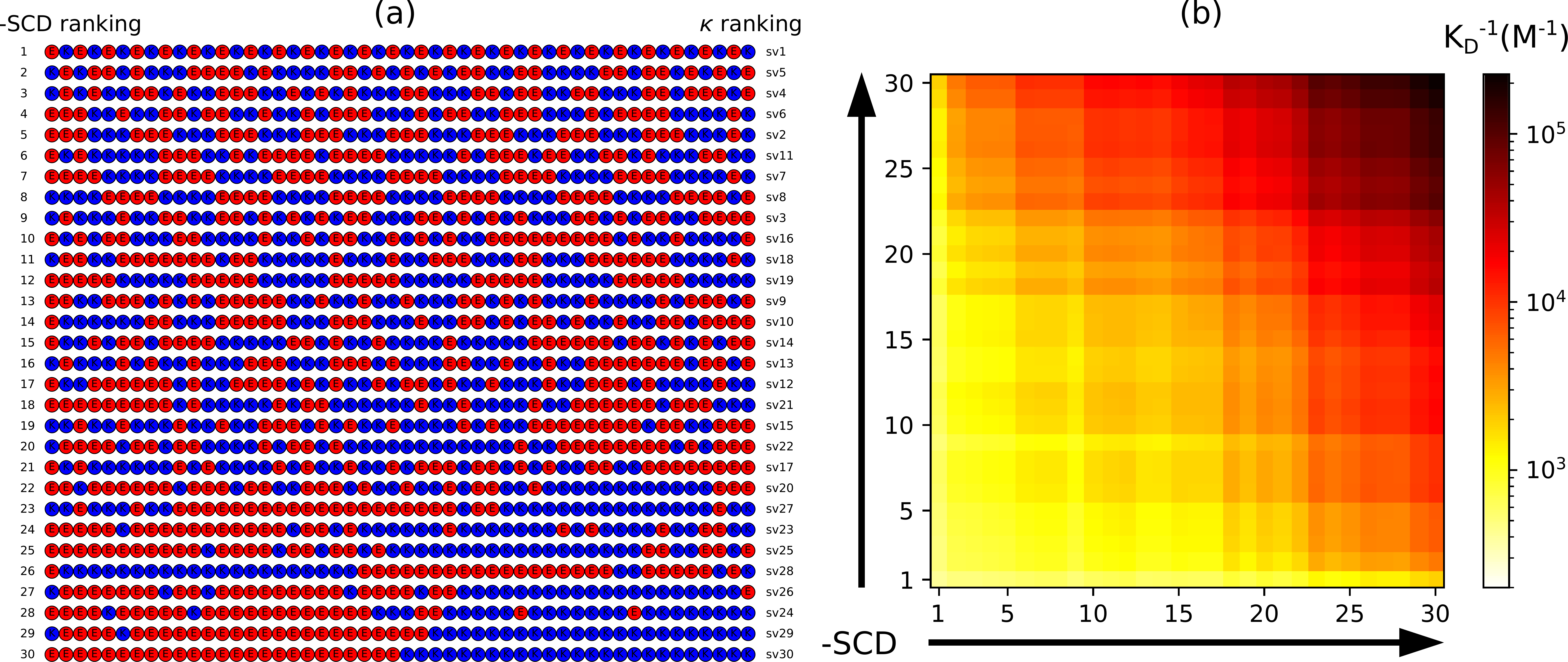}
    \caption{Fuzzy complex binding depends strongly on sequence charge patterns.
(a) The 30 sv sequences (red: $-1$, blue: $+1$) ordered by their SCD 
values (left) whereas the number after the ``sv" (right) indicates their 
ranking by $\kappa$ (Ref.~\citen{Das2013}). (b) Heatmap of 
binding affinities of all $30\times30$ sv pairs. Sequences with 
higher $-$SCD values bind more tightly.}
    \label{fig:KD_sv_seqs}
\end{figure}

\begin{figure}
    \centering
    \includegraphics[width=0.4\columnwidth]{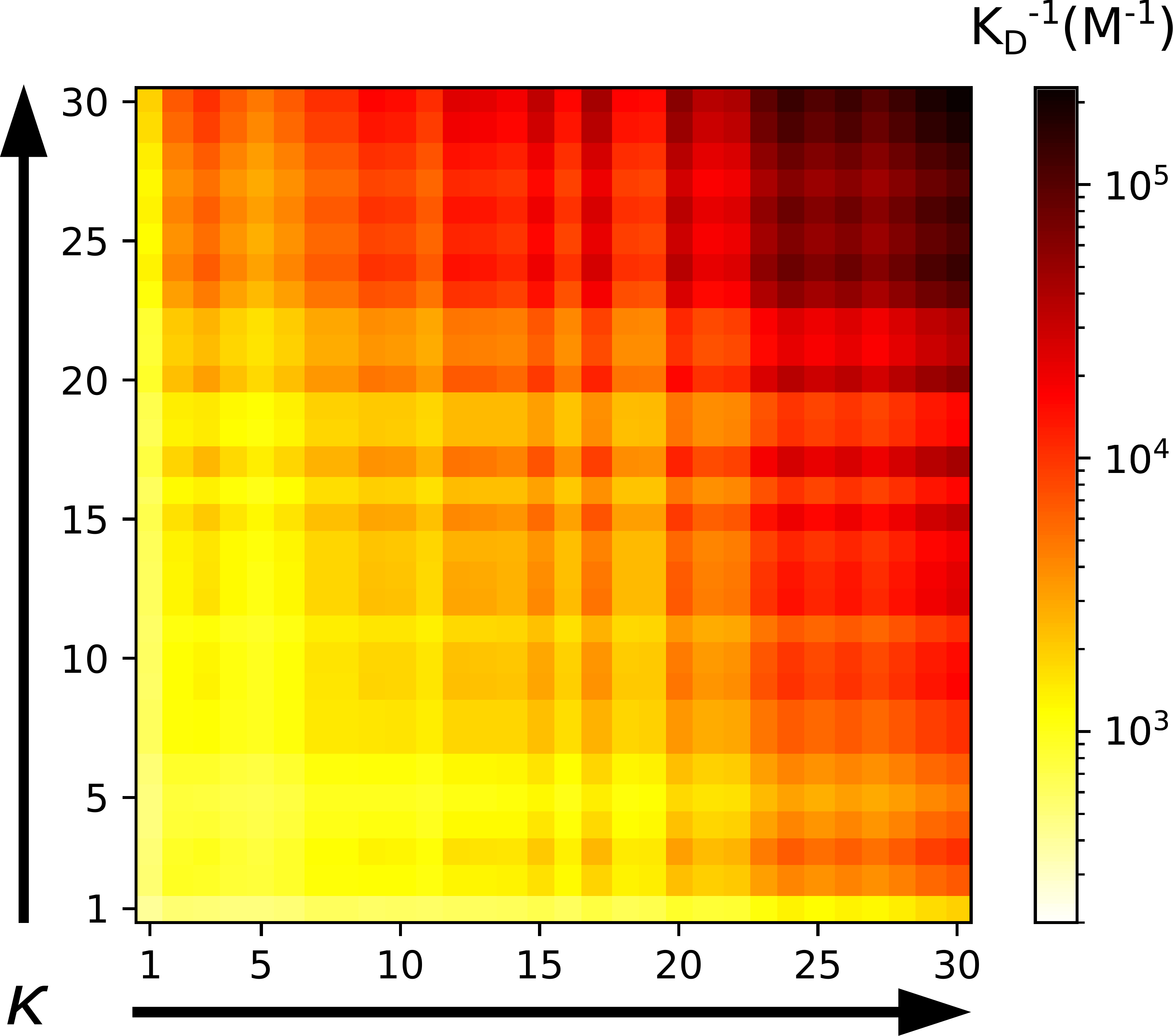}
    \caption{Heatmap of binding affinities of all 30 $\times$ 30 pairs of
overall charge neutral sv sequences
arranged in increasing value of the $\kappa$
parameter of Das and Pappu\cite{Das2013} along both axes. Consistent
with the trend shown in Fig.~\ref{fig:KD_sv_seqs}b for SCD dependence,
sequences with higher $\kappa$ values here are seen to have generally
higher binding affinities.}
    \label{fig:KD_kappa}
\end{figure}

We apply Eq.~\ref{eq:B2jSCD} to the set of 30 fully charged, overall 
neutral 50-monomer sv sequences introduced by Das and Pappu. Each
of these sequences contains 25 positive ($+$) and 25
negative ($-$) charges but they have different charge patterns\cite{Das2013} 
as quantitied by $\kappa$ and SCD~\cite{Lin2017b} (Fig.~\ref{fig:KD_sv_seqs}a).
Different binding constants ($\KD^{-1}$) ranging widely from 
under 5 $\mu$M to over 2 mM are predicted by Eq.~\ref{eq:B2jSCD} 
for the 900 sv sequence pairs, exhibiting a general trend of increasing
binding affinity with increasing charge segregation of the interacting IDPs
as measured by SCD (Fig.~\ref{fig:KD_sv_seqs}b) and $\kappa$
(Fig.~\ref{fig:KD_kappa}).


{\bf\large New analytical relationship with phase separation.}
jSCD characterizes not only binary fuzzy IDP complexes but also IDP phase
separation. In the random phase approximation (RPA) theory of phase
separation~\cite{Lin2016,Lin2017a} of overall charge neutral sequences
in the absence of salt and short-range cutoff of Coulomb interaction
(Eqs.~39 and 40 of Ref.~\citen{Lin2017a} with ${\tilde k}^2[1+{\tilde k}^2]
\to {\tilde k}^2$), the electrostatic free energy $\fel$ may
be expanded through $O(\lb^2)$ as
\begin{equation}
	\begin{aligned}
	\fel &=\int_0^{\infty}\frac{dkk^2}{4\pi^2}
	    \left\{\ln\left[1+\frac {4\pi\phi_m}{k^2T^*N}\langle 
\ch{}|\GaussianPM{}{k}|\ch{}\rangle\right]-\frac {4\pi\phi_m}{k^2T^*N}\langle 
\ch{}|\GaussianPM{}{k}|\ch{}\rangle\right\} \\
	&=-\frac{2\phi_m^2}{T^{*2}N^2}\int_0^{\infty}\frac{dk}{k^2}\langle
	    \ch{}|\GaussianPM{}{k}|\ch{}\rangle^2+O( \lb^3)\\
	&=-\frac{\phi_m^2}{T^{*2}}\sqrt{\frac {8\pi} 3}  \jSCD(\ch{},\ch{})
	    +O( \lb^3),
	\end{aligned}
	\label{eq:RPAeqn}
\end{equation}
where $N$ is chain length and $\phi_m$ is volume fraction of the IDP,
$T^* \equiv b/\lb$ is reduced temperature, and 
$\bra{\ch{}}\GaussianPM{}{k}\ket{\ch{}} = 
\sum_{s,t=1}^N \ch{}_s\ch{}_t\exp(-k^2|s-t|/6)$ is the charge structure 
factor ($\sum_{s=1}^N\sigma_s=0$ for neutral sequences). 
The $\phi_m^2$ term in Eq.~\ref{eq:RPAeqn} 
allows for an approximate sequence-dependent
Flory-Huggin (FH) theory of phase separation, which we term jSCD-FH, 
with an effective FH $\chi$ parameter 
\begin{equation}
\chi(\ch{},\ch{}) \equiv \sqrt{\frac {8\pi}3}
\frac{\jSCD(\ch{},\ch{})}{T^{*2}}
\; .
    \label{eq:chieff}
\end{equation}
For two IDP species $A,B$, one similarly obtains
$\chi(\ch{A},\ch{B})=\sqrt{8\pi/3}[\jSCD(\ch{A},\ch{B})]/T^{*2}$ and
\begin{equation}
\fel = 
    -\chi(\ch{A},\ch{A})\phi_A^2 
    -2\chi(\ch{A},\ch{B})\phi_A\phi_B
    -\chi(\ch{B},\ch{B})\phi_B^2
    +O( \lb^3)
\; 
    \label{eq:RPAeqn2}
\end{equation}
in the form of the FH interaction terms for
a-two-IDP species system (Eq.~27 of Ref.~\citen{Lin2017c}). 

\begin{figure}[ht]
    \centering
    \includegraphics[width=0.65\columnwidth]{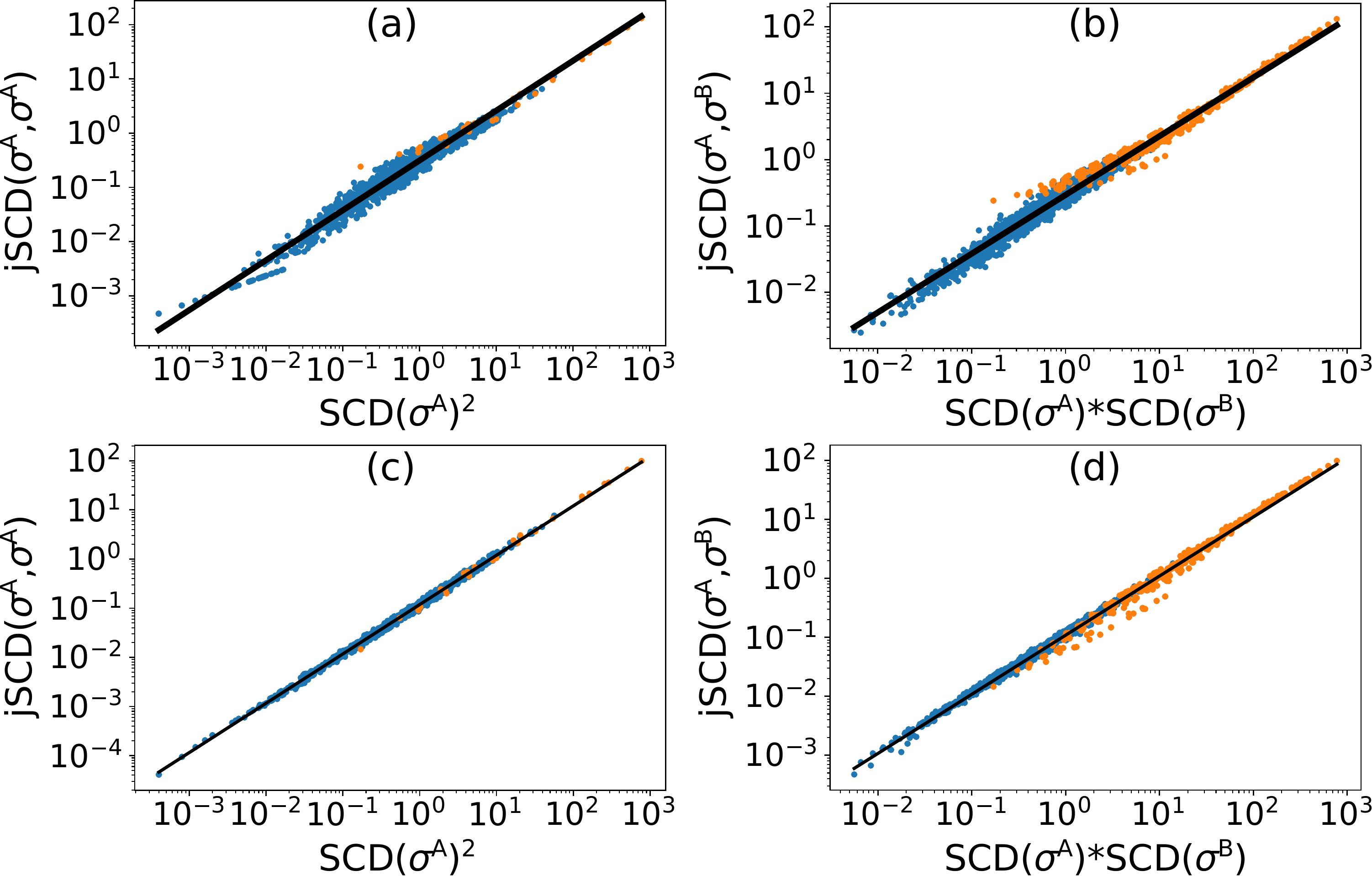}
    \caption{Correlation between single- and double-chain sequence charge
pattern parameters. $\jSCD$ vs $\SCD^2$ scatter plots 
for homotypic (a, c) or heterotypic (b, d) pairs
among the $30\times 30$ pairs of sv sequences (orange) and
1,000 random pairs of partially-charged, overall neutral 
50mers (blue) interacting via a pure Coulomb 
potential (a, b) or a Coulomb potential with a short-range cutoff (c, d).
Black lines are power-law regressions; square of Pearson 
coefficient $r^2=$ (a) 0.983, (b) 0.967, (c) 0.997, and (d) 0.994.
The correlation is good for both jSCD and jSCD$_{\rm cutoff}$ but
their fitted scaling exponents are not identical.
Apparently, SCD $<0$ for all overall charge neutral sequences
(see discussion in Supporting Information and Fig.~\ref{fig:eigens}).
}
    \label{fig:jSCDvsSCD}
\end{figure}

\begin{figure}[H]
    \centering
    \includegraphics[width=0.40\columnwidth]{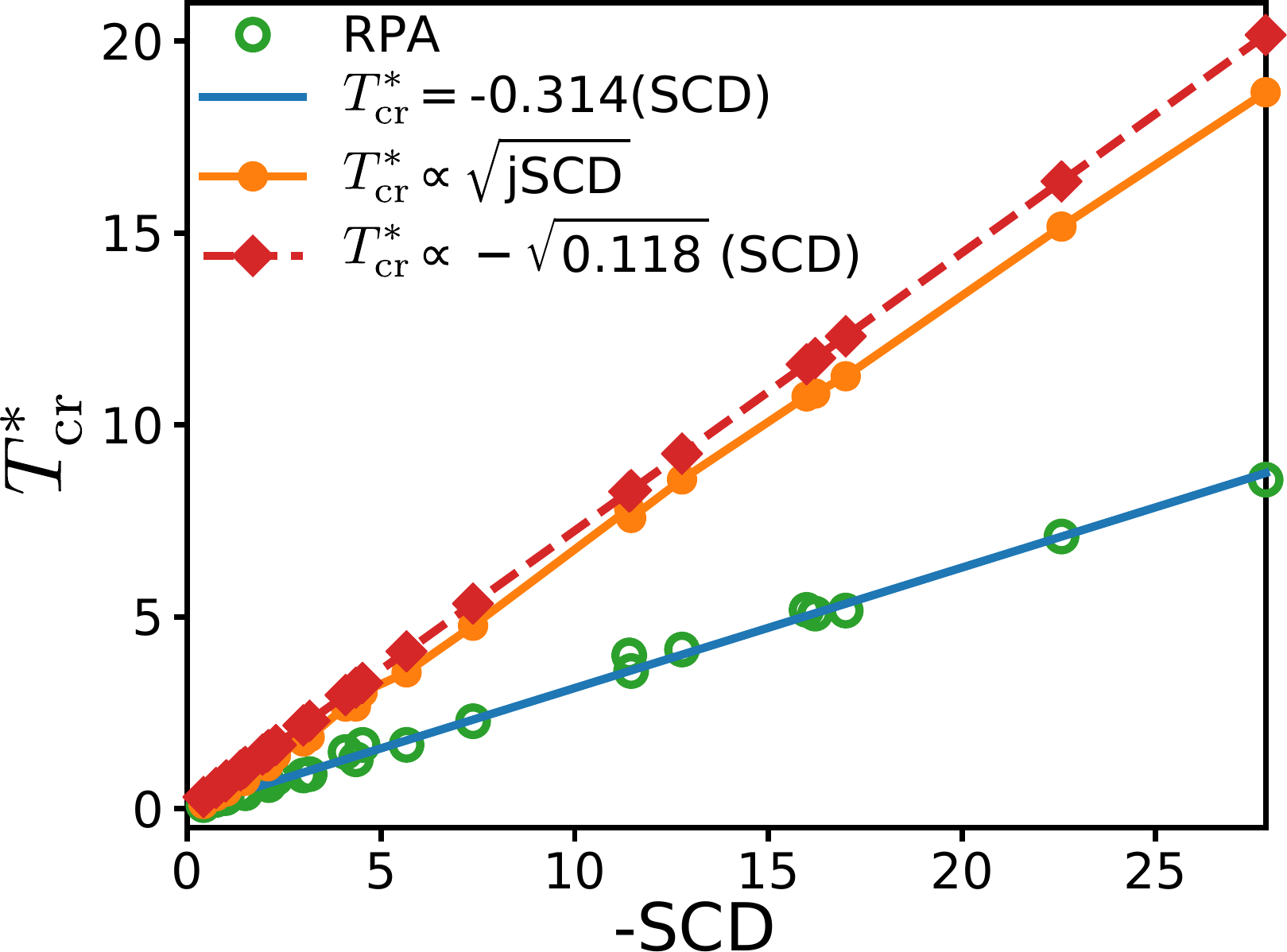}
    \caption{Approximate mean-field jSCD-FH phase separation theories
entail stronger dependence of critical temperature $T^*_{\rm cr}$ on
SCD than that predicted by RPA theory. Results shown are for the
30 sv sequences of Das and Pappu~\cite{Das2013}. Critical temperatures
calculated using RPA (green symbols) and its linear fit
$T^*_{\rm cr} =-0.314\times\SCD$ (blue line) are taken from Fig.~3b
of Ref.~\citen{Lin2017b}.
$T^*_{\rm cr}$ values computed here based on the jSCD-FH result in
Eq.~\ref{eq:Tc_jSCD} and the jSCD$_{\rm cutoff}$ expression in
Eq.~\ref{eq:jSCDcutoff}, i.e.,
$T^*_{\rm cr} = 2.11\sqrt{\jSCD_{\rm cutoff}}$,
are plotted in orange. The linear fit to the data points is provided
in the same color.
Slightly different jSCD-FH $T^*_{\rm cr}$ values are obtained
using the formula $T^*_{\rm cr} = -2.11\sqrt{0.118}\times\SCD$ solely by
replacing the actual $\jSCD_{\rm cutoff}$ values with the fitted value
$\jSCD_{\rm cutoff}=0.118\times(\SCD)^2$ deduced from
Fig.~\ref{fig:jSCDvsSCD}c.
Data in this plot indicate that both of the two jSCD-FH formulations
capture the $T^*_{\rm cr}\; \tilde{\propto}\;\SCD$
relation~\cite{Lin2017b} very well
but overestimate the phase separation propensities relative to the
RPA-predicted propensities for all 30 sv sequences.}
\label{fig:T_SCD_3form}
\end{figure}

\begin{figure}[H]
    \centering
    \includegraphics[width=0.55\columnwidth]{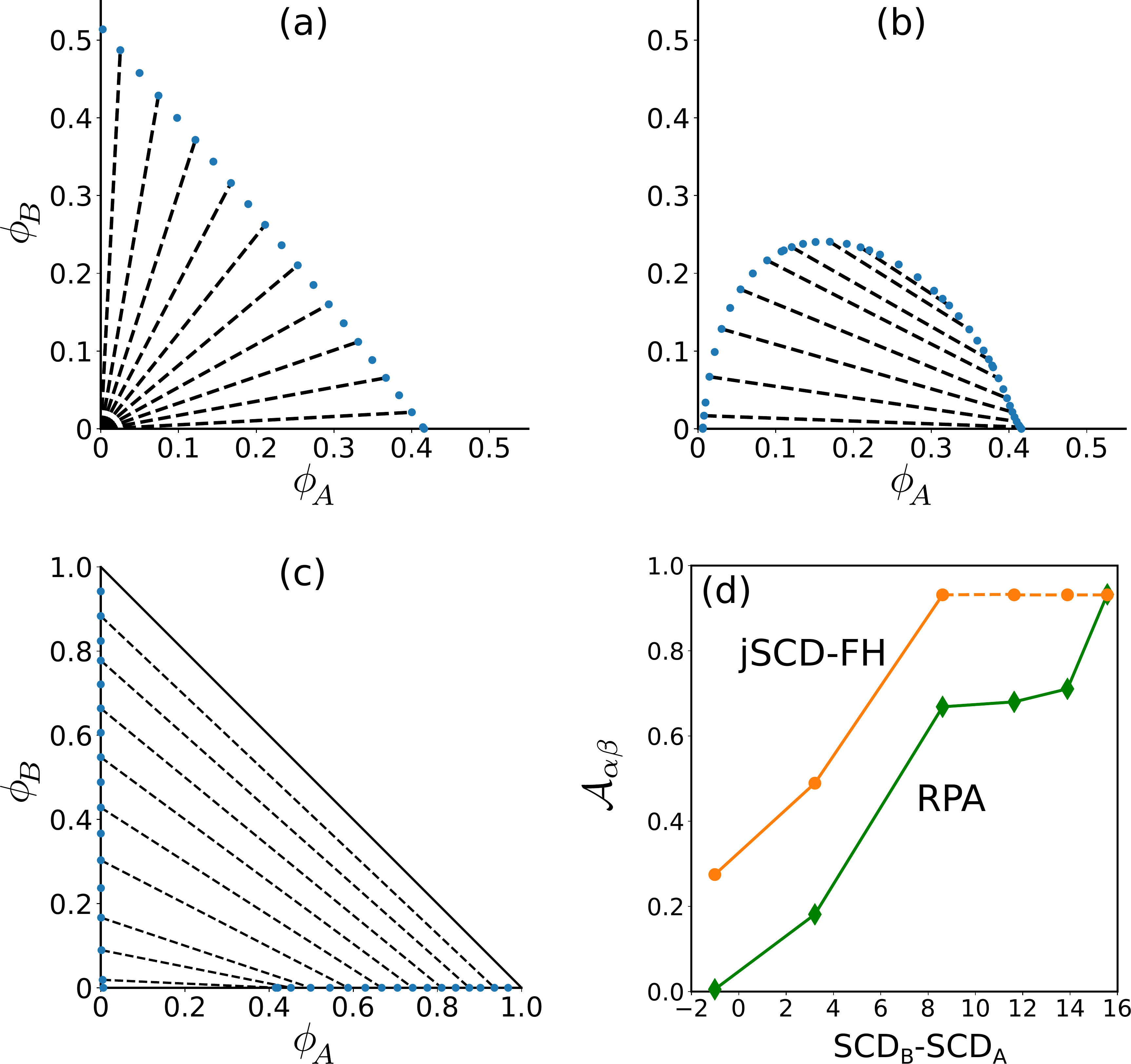}
    \caption{Binary phase diagrams generated by the approximate
jSCD-based effective Flory-Huggins
(jSCD-FH) interaction free energy given by Eq.~\ref{eq:RPAeqn2} 
with the $\chi$ parameters given by Eq.~\ref{eq:chieff} 
with $T^*=10$. Sequence $A$ is sv28 and sequence $B$s are
(a) sv24, (b) sv25, and (c) sv20. $\phi$s are volume fractions of the
polyampholytes.
Blue dots are numerically solved phase-separated states
$\alpha\equiv(\phi_A^\alpha,\phi_B^\alpha)$ and
$\beta\equiv(\phi_A^\beta,\phi_B^\beta)$ [the $\beta$ here for labeling
phase-separated
states is not to be confused with the reciprocal Boltzmann factor 
$1/k_{\rm B}T$]; black dashed lines are tie lines
connecting an $\alpha$--$\beta$ pair of coexisting states.
Consistent with the RPA phase diagrams provided in Fig.~3 of
Ref.~\citen{Lin2017c}, panels (a)--(c) here of jSCD-FH results show
the same general trend that sequences with similar $\SCD$s coalesce whereas
those with significantly different $\SCD$s exclude each other; but
the degree of exclusion predicted by the present jSCD-FH theory is significantly
higher than that predicted previously by RPA theory.
(d) Variation of the composition asymmetry measure, ${\cal A}_{\alpha\beta}$,
which is a demixing parameter (vertical axis), with the difference in SCD
values of the sequence pair (horizontal axis; $\SCD_A=\SCD(\ch{A})$, 
$\SCD_B=\SCD(\ch{B})$). The measure
${\cal A}_{\alpha\beta} \equiv (2/\pi)\langle
|\tan^{-1}(\phi_A^\alpha/\phi_B^\alpha) -\tan^{-1}(\phi_A^\beta/\phi_B^\beta)|
\rangle$, where the $\langle\cdots\rangle$ average is over
all tie-line connected $\alpha$--$\beta$ pairs,
is defined in Eq.~26 of Ref.~\citen{Lin2017c} to quantify the
tendency of two sequences $A$ and $B$ in a solution system
to demix upon separation into two phases $\alpha$ and $\beta$.
The orange jSCD-FH data points here are seen to be always higher
than the corresponding green RPA data points, indicating that the more
approximate mean-field jSCD-FH formulation
always overestimates demixing propensity. Lines joining data points are
guides for the eye. The last three jSCD-FH data points are connected by
dashed lines instead of solid lines to underscore the fact that
${\cal A}_{\alpha\beta}$ is already saturated at the third
(sv28--sv20) sequence pairs shown and the remaining ${\cal A}_{\alpha\beta}$
data points for larger {\hbox{SCD$_B$ $-$ SCD$_A$}} differences remain
at the same saturated value.}
    \label{fig:FH_RPAexpand}
\end{figure}

Recognizing $\chi=\chi_{\rm cr}=(\sqrt{N}+1)^2/(2N)$ at the FH critical
temperature $T^*_{\rm cr}$, Eq.~\ref{eq:chieff} suggests that for $N=50$, 
\begin{equation}
T^*_{\rm cr}(\ch{}) \approx 2.11 \times \jSCD^{1/2}(\ch{},\ch{})
\; .
	\label{eq:Tc_jSCD}
\end{equation}
A strong correlation between $\jSCD$ and the product of its two component
$\SCD$s is suggested by Fig.~\ref{fig:KD_sv_seqs}b. Indeed,
for the 30 sv sequences as well as 1,000 randomly generated overall charge
neutral 50mer sequences (see Supporting Information for description),
$\jSCD(\ch{},\ch{}) \approx 0.293\times |\SCD(\ch{})|^{1.77}$ and 
$\jSCD(\ch{A},\ch{B}) \approx 0.313[\SCD(\ch{A})\times\SCD(\ch{B})]^{0.920}$
(Fig.~\ref{fig:jSCDvsSCD}a,b). The correlations are excellent aside
from slightly more scatter around SCD$^2\sim 1$.
To assess the robustness of these correlations, we consider also
a modified Coulomb potential $\lb[1-\exp(-r/b)]/r$ with short-range 
cutoff used in 
RPA~\cite{Ermoshkin2003, Lin2016,Lin2017a,Lin2017b,Lin2017c}
to derive a modified $\jSCD$,
\begin{equation}
\jSCD_{\text{cutoff}}(\ch{A},\ch{B}) \equiv\frac{1}{\NA\NB}
\sqrt{\frac 3 {2\pi}}\int_0^{\infty}\frac{dk}{k^2(1+k^2)^2}
    \sum_{s,t=1}^\NA\sum_{l,m=1}^\NB
        \ch{A}_s\ch{A}_t\ch{B}_l\ch{B}_m
        e^{-\frac{1}{6}(kb)^2[|s-t|+|l-m|]}
\; ,
    \label{eq:jSCDcutoff}
\end{equation}
and find that 
$\jSCD_{\rm cutoff}(\ch{},\ch{})\approx 0.118\times |\SCD(\ch{})|^{2.007}$ and
$\jSCD_{\rm cutoff}(\ch{A},\ch{B})\approx 0.109[\SCD(\ch{A})
\times\SCD(\ch{B})]^{1.003}$ 
(Fig.~\ref{fig:jSCDvsSCD}c,d).
Interestingly, combining the $\jSCD_{\rm cutoff}(\ch{},\ch{})$ scaling and
Eq.~\ref{eq:Tc_jSCD} rationalizes the 
$T^*_{\rm cr}\; \tilde{\propto}\;\SCD$ scaling in Ref.~\citen{Lin2017b}
(Fig.~\ref{fig:T_SCD_3form}); and
this analytical result is in line with the relation between $\btwo$ and
$T^*_{\rm cr}$ deduced from explicit-chain simulations~\cite{Dignon2018}.
Taking into account also the $\jSCD_{\rm cutoff}(\ch{A},\ch{B})$ 
scaling and Eq.~\ref{eq:RPAeqn2} rationalizes the 
$\chi(\ch{A},\ch{B})= \sqrt{\chi(\ch{A},\ch{A})\;\chi(\ch{B},\ch{B})}$
relation in Ref.~\citen{Lin2017c} (Fig.~\ref{fig:FH_RPAexpand}). 
Not unexpectedly, in both cases, approximate mean-field jSCD-FH produces a trend
consistent with RPA, but entails a sharper dependence of phase behaviors 
on SCD than that predicted by RPA (Figs.~\ref{fig:T_SCD_3form} and 
\ref{fig:FH_RPAexpand}).
In this connection, it is instructive to note that the general trend 
of sequence dependent critical temperature of polyampholyte phase 
separation has recently been shown to agree largely with that obtained 
from field-theoretic simulations~\cite{Danielsen2019},
despite the RPA's expected limitations 
in accounting for polyampholyte 
phase behaviors at very low concentrations.

Previously, the tendency of the populations of two polyampholytes $A$ and $B$ 
to demix upon phase separation (as quantified, e.g., 
by an ${\cal A}_{\alpha\beta}$
parameter) was reported to correlate with their SCD difference 
$\SCD(\ch{B})-\SCD(\ch{A})$
(Ref.~\citen{Lin2017c} and Fig.~\ref{fig:FH_RPAexpand}d).
In view of the above theoretical development and the fact that 
${\cal A}_{\alpha\beta}$ $\sim$ $\SCD(\ch{B})-\SCD(\ch{A})$
was observed only for a set of six sv pairs 
($A$ and $B$) all having sv28 as sequence $A$, this previously
observed empirical correlation
should now be viewed as a special case of an expected general correlation
between $\jSCD(\ch{A},\ch{B})$ and the tendency for demixing of 
sequences $A$ and $B$ upon phase separation because 
in the special case when $\SCD(\ch{A}) = {\rm constant}$,
$\jSCD(\ch{A},\ch{B})$ $\tilde{\propto}$ $\SCD(\ch{A})\times \SCD(\ch{B})$
$\tilde{\propto}$ $[\SCD(\ch{B})-\SCD(\ch{A})] + {\rm constant}$.
%


\begin{figure}[t]
    \centering
    \includegraphics[width=\columnwidth]{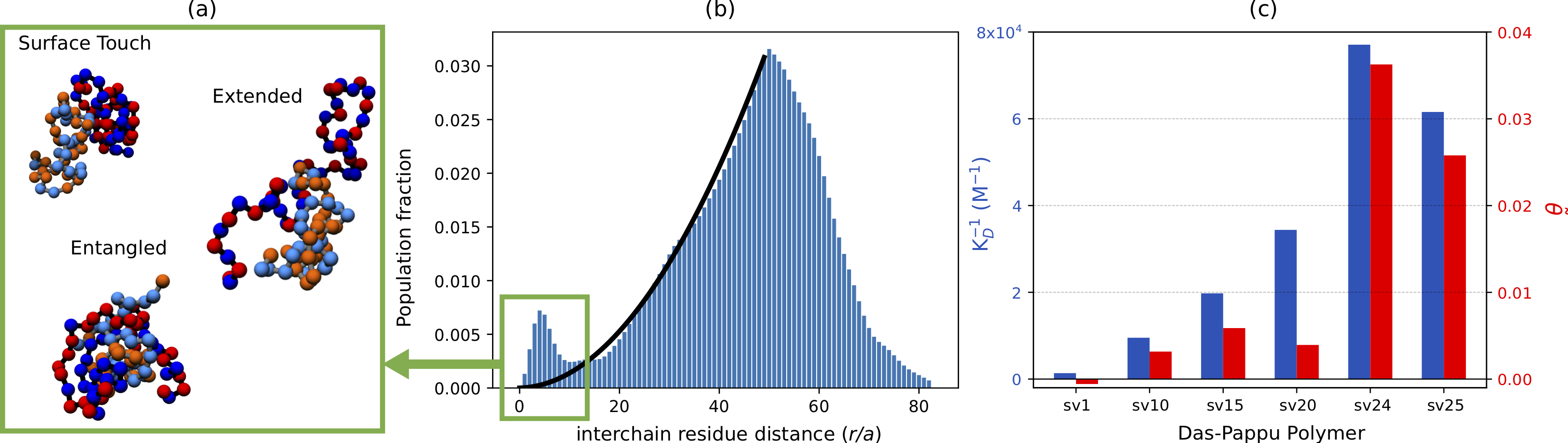}
    \caption{Comparing analytical theory against explicit-chain 
simulation. Results from simulations are for $T^*=0.35$. 
(a) Snapshots of fuzzy complexes of sv28 
(cyan/orange: $+/-$) with different partners (blue/red: $+/-$): sv24 
(surface touch), s25 (entangled) and sv1 (extended).
    (b) Distribution (histogram) of sv24--sv28 interchain residue-residue 
distance among 1,000,000 snapshots. The small-$r$ peak region
(marked by the green frame) is the bound state. The
black curve is the baseline distribution of distance 
between a pair of non-interacting particles in the same simulation box.
    (c) Theoretical $\KD$ (blue) vs simulated $\tilde{\theta}$ (red),
of sv28 with various partners (horizontal axis),
where $\tilde{\theta} \equiv \theta - \theta_0$
with $\theta_0 = 4\pi\times 10^3/(3\times100^3)$ being the 
baseline probability that two non-interacting particles is 
$< 10a$ apart. The $\tilde{\theta}<0$ result for sv1 means that the net 
interaction between sv28 and sv1 is repulsive in the simulation.}
    \label{fig:Sim}
\end{figure}

{\bf\large Theory-predicted trend is consistent with simulations and
Kuhn length renormalization}.
We now assess our approximate theory by comparing its predictions with 
coarse-grained molecular dynamics simulations~\cite{Das2018b} of six sv 
sequence pairs.  Details of the explicit-chain model is in the Supporting 
Information. Because bound IDPs in a fuzzy complex are dynamic,
their configurations are diverse. The IDP chains in
some bound configurations are relatively open, some are highly intertwined, 
others can take the form of two relatively compact chains interacting
favorably mostly via residues situated on the surface of their individually
compact conformations (Fig.~\ref{fig:Sim}a). Taking into account this
diversity, we sample all intermolecular residue-residue distances between 
the model IDPs (rather than merely their center-of-mass distances) and
use the appearance of a bimodal distribution to define binding 
(Fig.~\ref{fig:Sim}b) with binding probability $\theta$ given by the
fractional area covered by the small-distance peak. To better quantify the 
role of favorable interchain interaction---rather than random collision---in
the formation of IDP complexes, we subtract a reference probability, 
$4\pi r_{\rm cut}^3/(3V)$, that two particles in a simulation box of size $V$
will be within the cutoff distance $r_{\rm cut}$ that defines the 
the small-distance peak in Fig.~\ref{fig:Sim}b;
and compare $\tilde{\theta} \equiv \theta - 4\pi r_{\rm cut}^3/(3V)$
with theoretical predictions.

For the sequence pairs considered, theoretical $\KD^{-1}$
is generally substantially higher than simulated $\KD^{-1}$ at the 
same temperature. The mismatch likely arises from differences in
the two models; for example, excluded volume is considered in the simulation
but not in the present analytical theory. For the same reason, a similar 
mismatch between theory and explicit-chain simulation has been noted in 
the study of phase separation of sv aequences~\cite{Lin2017b, Das2018b}.
Nonetheless, sequence-dependent trends of binding predicted by theory 
and simulation are largely similar (Fig.~\ref{fig:Sim}c). Notably,
both theory and simulation posit that sv24--sv28 binds more strongly
than sv25--sv28, exhibiting a rank order that is consistent with SCD (sv24 has
a larger $-$SCD value than sv25) but not $\kappa$ (sv24 has 
a smaller $\kappa$ parameter than sv25).

However, theory and simulation disagree on the rank order of sv15--sv28 
and sv20--sv28 binding affinities (Fig.~\ref{fig:Sim}c).
As a first step in addressing this discrepancy, we examine more closely
the impact of using a Gaussian-chain assumption to derive the $\btwo$
formula in Eq.~\ref{eq:B2_F1F2}.


%
The Gaussian-chain approximation in the general 
formula for $\btwo$ in Eq.~\ref{eq:B2_general} is for tractability. But
in reality intrachain residue-residue correlation is physically affected by
intrachain Coulomb interaction, as illustrated by the simulation snapshots
in Fig.~\ref{fig:Sim}a. Several analytical approaches have been proposed
to account for this effect approximately, including that of Sawle
and Ghosh for polyampholytes~\cite{Sawle2015} and 
that of Shen and Wang for polyelectrolytes~\cite{ShenWang2018}.
Here we focus on the method in Ref.~\citen{Sawle2015}, which entails 
deriving sequence-dependent effective, or 
renormalized, Kuhn lengths, denoted as $x^i_{st}b$ for residue pair $s,t$
in chain $i$, to replace the ``bare'' Kuhn length $b$ in the original 
simple Gaussian formulation. In other words, 
the modification
\begin{equation}
\left[\hat{P}^i(\kk)\right]_{st} 
\approx
\exp \left[ -\frac{(kb)^2}{6}|s-t| \right]
\to
\exp \left[ -\frac{(kb)^2}{6}x^i_{st}|s-t| \right]
\;
\end{equation}
is applied to Eq.~\ref{eq:Gaussian}. 
In this approach, instead of assuming that the
conformational distribution of each of the two IDP chains in our binary
interacting IDP-IDP system is that of a simple Gaussian chain as if it
experiences no interaction other than the contraints of chain connectivity,
the impact of the {\it intrachain} part of the interaction in the system
on the conformational distribution of an individual IDP chain is taken
into account approximately by treating the IDP as a modified Gaussian
chain with a renormalized Kuhn length~\cite{Sawle2015}.
As such, it should be noted that this renormalization procedure is 
performed on a single isolated chain without addressing effects of 
interchain interactions.

Recognizing that the simple Gaussian-chain correlation function in
Eq.~\ref{eq:Gaussian} 
is a consequence of a single-chain Hamiltonian $\HH{i}_0[\R^i]$ containing
only terms for elastic chain connectivity, viz.,
\begin{equation}
\HH{i}_0[\R^i] = \frac{3}{2b^2} \sum_{s=1}^{N_i-1}\left\vert \R^i_{s+1} -
\R^i_{s} \right\vert^2
\; ,
\label{eq:H0}
\end{equation}
we now also take into consideration an intrachain interaction potential
$\U^i[\R^i]$ that includes electrostatic interaction and excluded-volume
repulsion,
\begin{equation}
\U^i[\R^i] = \sum_{s>t=1}^{N_i}\left[
 \lb\ch{i}_s\ch{i}_t \frac{e^{-\kappaD |\R^i_s-\R^i_t|}}{|\R^i_s-\R^i_t|} +
        w^i\delta^3(\R^i_s-\R^i_t)
        \right]
\label{eq:U}
\; ,
\end{equation}
where $w^i$ is the two-body excluded-volume repulsion strength for chain $i$.
For the 30 sv sequences~\cite{Das2013} used in the present
analysis, the $w^i$ values obtained from matching theory with result from
explicit-chain atomic simulation conducted in the ``intrinsic
solvation'' limit in the absence of electrostatic interactions\cite{Das2013}
are available from Table~1 of Sawle and Ghosh~\cite{Sawle2015}.
A full Hamiltonian $\HH{i}$ is then given by the sum of
Eqs.~\ref{eq:H0} and \ref{eq:U}:
\begin{equation}
\HH{i}[\R^i] =  \HH{i}_0[\R^i]  + \U^i[\R^i]
\; .
\end{equation}
We assume, as in Ref.~\citen{Sawle2015}, that the full Hamiltonian
can be approximated as the Hamiltonian $\T_x^i[\R^i]$ for a modified
Gaussian chain with an effective Kuhn length $x^i b$, which is
equivalent to $N_i b^2\rightarrow N_ib(x^i b)$ while holding
the total contour length $N_ib$ unchanged (cf. Eqs.~1 and 2
of Ref.~\citen{Sawle2015}). In other words~\cite{Kings},
\begin{equation}
\HH{i}[\R^i] \approx  \T_x^i[\R^i] \equiv
\frac{3}{2xb^2} \sum_{s=1}^{N_i-1}\left\vert\R^i_{s+1} - \R^i_{s}\right\vert^2
\; ,
\label{eq:Tx}
\end{equation}
where $x$ is to be determined by the variational approach described
in Ref.~\citen{Sawle2015}.
Here we briefly summarize the concept and result, and refer the readers
to the original paper\cite{Sawle2015} for methodological details. The
approach consists of expressing the full Hamiltonian as a sum of the
``principal" $\T_x^i$ component and a ``perturbative'' $\Delta\HH{i}_x$ term:
\begin{equation}
\HH{i}[\R^i]  =  \T_x^i[\R^i] + \Delta\HH{i}_x[\R^i],
\end{equation}
where, by Eqs.~\ref{eq:H0}, \ref{eq:U} and \ref{eq:Tx},
\begin{equation}
\Delta\HH{i}_x[\R^i] =  \frac{3}{2b^2}\left( 1-\frac{1}{x}\right)
\sum_{s=1}^{N_i-1}\left\vert \R^i_{s+1} - \R^i_{s} \right\vert^2
        +  \sum_{s>t=1}^{N_i}\left[
     \lb\ch{i}_s\ch{i}_t \frac{e^{-\kappaD |\R^i_s-\R^i_t|}}{|\R^i_s-\R^i_t|} +
        w^i\delta^3(\R^i_s-\R^i_t)\right]
\; .
\end{equation}
Making use of the form in Eq.~\ref{eq:Tx}, the full thermodynamic average
$\Avg{A}$---Boltzmann-weighted by the full Hamiltonian $\HH{i}$---of any
physical observable $A$ can be cast as an expansion in
the power of the perturbative Hamiltonian $\Delta\HH{i}_x$
(Eq.~3 of Ref.~\citen{Sawle2015}):
\begin{equation}
\Avg{A} = \Avg{A}_x + \Avg{A}_x\Avg{\Delta\HH{i}_x}_x -
\Avg{ A\Delta\HH{i}_x }_x + O\left[(\Delta\HH{i}_x)^2\right]
\; ,
        \label{eq:A_var}
\end{equation}
where the averages $\Avg{\cdots}$ and $\Avg{\cdots}_x$ are defined by
\begin{subequations}
\begin{align}
\Avg{A} \equiv &
        \; \frac{\int \DD[\R^i] A[\R^i] e^{-\HH{i}[\R^i]}}{\int \DD[\R^i]
e^{-\HH{i}[\R^i]}} \; , \\
\Avg{A}_x \equiv &
        \; \frac{\int \DD[\R^i] A[\R^i] e^{-\T^i_x[\R^i]}}{\int \DD[\R^i]
e^{-\T^i_x[\R^i]}}
\; .
\end{align}
\end{subequations}
For any observable $A$ of interest, an optimal $x$ in this
formalism is obtained by minimizing the difference between the
averages weighted by the full $\HH{i}$ and the
approximate $\T^i_x$ through eliminating the $O(\Delta\HH{i}_x)$
term in Eq.~\ref{eq:A_var}. Imposing this condition allows us to solve
for an optimal set of $x^i_{st}$ 
for a given $A$. Comparisons by Ghosh and coworkers
of results from this theoretical approach against those from explicit-chain
simulations have demonstrated that this is a
rather accurate and effective method~\cite{Sawle2015,HuihuiGhosh2020}.
Ideally, the correlation functions $\left[\hat{P}^i(\kk)\right]_{st}$ 
themselves should be used as observables for the optimization; but that 
leads to insurmountable technical difficulties. Thus, following 
Ref.~\citen{Sawle2015} (Eq.~11 of this reference), we 
use $|\R^i_s-\R^i_t|^2$ as observables to optimize $x^i_{st}$s. 
Accordingly, for each residue pair $s^i,t^i$ on chain $i$, an optimized
$x$ factor, $x^i_{st}$, is obtained by solving the equation
\begin{equation}
\Avg{|\R^i_s-\R^i_t|^2}_{x^i_{st}}\Avg{\Delta\HH{i}_{x^i_{st}}}_{x^i_{st}}
= \Avg{ |\R^i_s-\R^i_t|^2\Delta\HH{i}_{x^i_{st}} }_{x^i_{st}}
\;
        \label{eq:R2_avg}
\end{equation}
using the formalism developed in Eqs.~6--10 of Ref.~\citen{Sawle2015}.
These solved $x^i_{st}$ values are then used to rescale the
two terms of the $X$ factor introduced in Eq.~\ref{eq:IX0}
to arrive at the expression
\begin{equation}
\btwo =
    \frac{4\pi\lb}{\kappaD^2}\nch{A} \nch{B}
    -4\lb^2 \int \frac{dk k^2}{(k^2+\kappaD^2)^2}
    \sum_{s,t=1}^\NA\sum_{l,m=1}^\NB
        \ch{A}_s\ch{A}_t\ch{B}_l\ch{B}_m
        e^{-\frac{1}{6}(kb)^2[x^A_{st}|s-t|+x^B_{lm}|l-m|]}
    \label{eq:B2_F1F2x}
\end{equation}
for the second virial coefficient in the
formulation with renormalized Kuhn lengths.
In the case of a salt-free solution of overall charge neutral polymers,
this expression reduces to
\begin{equation}
   \left. \btwo^{\rm eff}\right|_{\kappaD\to0,\nch{A}\nch{B}=0}
   =
   4\sqrt{\frac{\pi}{6}}\lb^2 b \sum_{s,t=1}^\NA\sum_{l,m=1}^\NB
    \ch{A}_s\ch{A}_t\ch{B}_l\ch{B}_m\sqrt{x^A_{st}|s-t|+x^B_{lm}|l-m|}
\; ,
    \label{eq:btwo_eG}
\end{equation}
which is the modified (renormalized) form of Eq.~\ref{eq:B2jSCD}.

\begin{figure}[t]
    \centering
    \includegraphics[width=0.9\columnwidth]{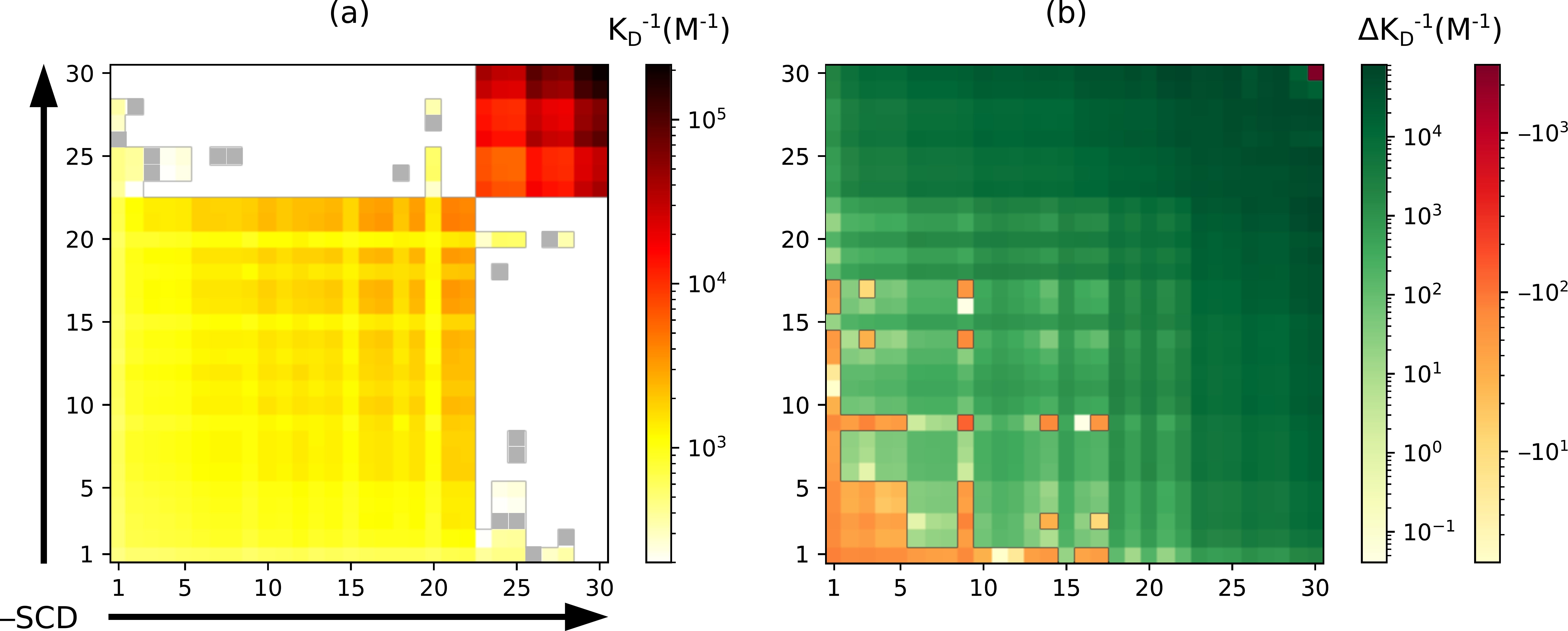}
    \caption{Heatmap of binding affinities of the $30\times 30$
overall charge neutral sv sequence
pairs computed using Eq.~\ref{eq:btwo_eG} in the formulation
with renormalized Kuhn lengths~\cite{Sawle2015}.
White squares indicate an unfavorable (repulsive) interaction and
grey squares indicate a weak $K_{\rm D}$ of greater than 5 mM.
The results are quite different from those provided in
Fig.~\ref{fig:KD_sv_seqs}b for the base theory with a bare
(not renormalized) Kuhn length. (b) Heatmap of difference in
the same sequence pairs' binding affinities predicted by the two theories
(base-theory prediction {\it minus} renormalized-Kuhn-lengths prediction).
In general, more charge segregated sequences, i.e., those with higher
$-$SCD values, exhibit a higher reduction in binding affinities when
intrachain interactions are accounted for approximately using
renormalized Kuhn lengths.}
\label{fig:KDeff}
\end{figure}

The resulting
heatmap of the $\KD$ values calculated in this manner
is provided in Fig.~\ref{fig:KDeff}a.
Unlike the results obtained using the base theory with a simple Gaussian
chain model (Fig.~\ref{fig:KD_sv_seqs}b), the theory of renormalized 
Kuhn lengths predicts that some sv sequence pairs do not bind at all,
as indicated by the white regions in Fig.~\ref{fig:KDeff}a.
Furthermore, instead of binding propensity being monotonic with charge
segregation (quantified by $-$SCD) as predicted by the base theory, some
sv sequence pairs deviate from the trend. Specifically,
highly charge segregated sequences with large $-$SCD values seem to
avoid interactions with sequences with only a medium charge segregation
with moderate $-$SCD values.


\begin{figure}[t]
    \centering
    \includegraphics[width=0.7\columnwidth]{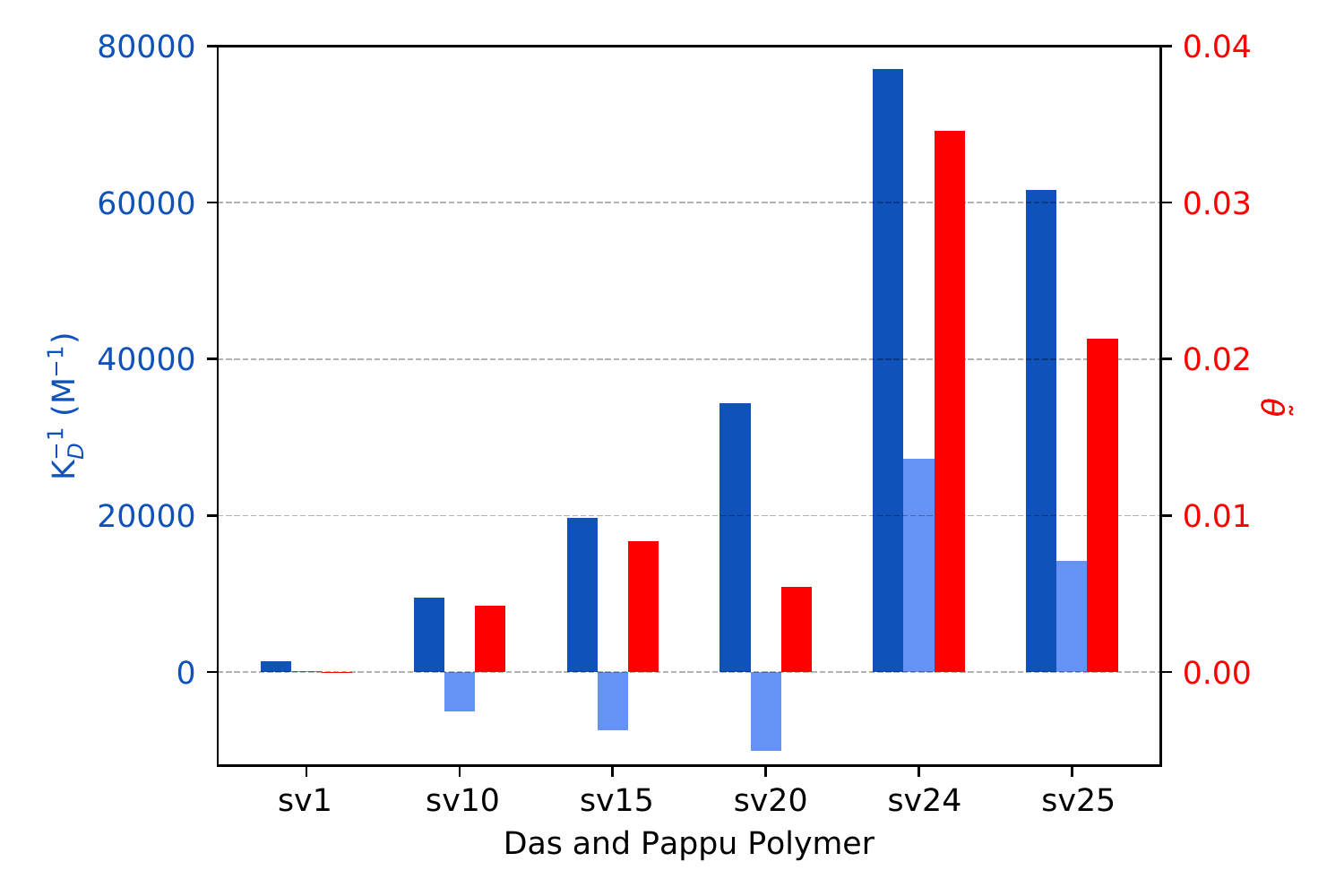}
    \caption{Binding affinities of example sv sequence pairs. Theoretical
and simulation results are provided for sv28 pairing individually with
sv1, sv10, sv15, sv20, sv24, and sv25. As in Fig.~\ref{fig:Sim},
predictions by theory using simple Gaussian chains
without renormalized Kuhn lengths (Eq.~\ref{eq:B2jSCD})
are shown in dark blue, explicit-chain simulation results, calculated
anew here using the regression method described in the Supporting
Information for $T^*=0.35$
(Eq.~\ref{eq:theta_regress} and Table~\ref{tb:sim-regress}),
are shown in red. Included here for comparison are predictions by theory using
renormalized Kuhn lengths, shown in light blue, as prescribed by
Eq.~\ref{eq:btwo_eG}.
It is noteworthy from this comparison that effects of intrachain
interactions on single-chain conformational distribution may
afford a partial rationalization for the discrepancy between
simple theory (dark blue) and explicit-chain simulation (red)
for sv20--sv28 binding but for the present case
such effects are likely overestimated by the method of renormalized Kuhn
lengths\cite{Sawle2015} to result in a net repulsion (negative light blue
bars for not only sv20--sv28 but also sv10--sv28 and sv15--sv28).}
\label{fig:KD_sim_all_reg}
\end{figure}

These contrasts between the base theory and the formulation with
renormalized Kuhn lengths are underscored in
Fig.~\ref{fig:KDeff}b where the numerical
differences in predicted binding affinities by the two formulations
are plotted.
Apparently, the approximate account of intrachain interactions afforded
by renormalized Kuhn lengths posits a larger decrease
in binding affinities relative to that predicted by the base theory
or high $-$SCD sequences than for low $-$SCD sequences.
The $\KD$s predicted by the two theories and the binding probabilities
obtained from explicit-chain simulations for several
example sv sequence pairs are compared in more detail in
Fig.~\ref{fig:KD_sim_all_reg}.
These predictions are physically intuitive as 
sequences with larger $-$SCDs generally
have stronger intrachain interactions, although the magnitude of the 
effect is likely overestimated.
With the last caveat, the higher simulated binding of sv15--sv28
relative to that of sv20--sv28 may be understood in terms of
sv20's more favorable intrachain interaction
(Fig.~\ref{fig:KD_sim_all_reg}). In this context,
it would be extremely interesting to explore in future investigations the 
impact of the improved formulation of $x^A_{st}$ and $x^B_{lm}$ 
proposed recently by Huihui and Ghosh~\cite{HuihuiGhosh2020} on
the association of sv model sequences and other polyampholytes.
In particular, for the polyelectrolyte H1-ProT$\alpha$ system considered 
above (Fig.~\ref{fig:Kd_H1_ProTa}), since the highly open individual
H1 and ProT$\alpha$ conformations at low salt are expected to entail 
more favorable H1-ProT$\alpha$ interactions than their less open individual 
conformations at high salt, an analytical theory 
with renormalized Kuhn lengths for individual IDP chains would likely lead to 
a higher salt sensitivity for $\KD$ and hence better agreement with 
experiments. This expectation, however, remains to be tested.


\section{Conclusions}

In summary, we have developed an analytical account of charge 
sequence-dependent fuzzy binary complexes with novel two-chain
charge pattern parameter jSCD emerging as a key determinant not only
of binary binding affinity but also of multiple-chain phase separation. 
The formulation elucidates the dominant role of conformational disorder
and sequence-specificity in IDP-IDP binding, and provides a 
footing for empirical correlation between single-
and two-chain IDP properties with their sequence-dependent phase-separation 
propensities~\cite{Lin2017b,Riback2017,Dignon2018,Aksimentiev2020,Rohit2020}.
While the formulation is limited inasmuch as it is a 
high-temperature approximation and further developments, including
extension to sequence patterns of uncharged 
residues~\cite{Pak2016,Wang2018,panag2020,Zheng2020},
are desirable, the charge sequence dependence predicted herein
is largely in line with explicit-chain simulation. As such, the
present formalism offers conceptual advances as well as utility
for experimental design and efficient screening of candidates 
of fuzzy complexes.

\section{Acknowledgements}

We thank Robert Best, Aritra Chowdhury, Julie Forman-Kay, Alex Holehouse,
Jeetain Mittal, Rohit Pappu, and Wenwei Zheng for helpful discussions, and 
Ben Schuler for insightful comments on an earlier version of this paper 
(arXiv:1910.11194v1) and sharing unpublished data. 
This work was supported by Canadian Institutes of Health 
Research grants MOP-84281, NJT-155930, Natural Sciences and Engineering 
Research Council of Canada Discovery grant RGPIN-2018-04351, and computational 
resources provided by Compute/Calcul Canada.


\ \

\noindent The authors declare no conflict of interest.

\newpage

\setcounter{equation}{0}
\setcounter{figure}{0}
\setcounter{table}{0}
\setcounter{page}{1}
\renewcommand{\theequation}{S\arabic{equation}} 
\renewcommand{\thefigure}{S\arabic{figure}} 
\renewcommand{\thetable}{S\arabic{table}} 
\renewcommand{\thepage}{S\arabic{page}} 



\noindent{\LARGE\titlefont Supporting Information}

\section{Derivation for $\btwo$ representations}

Starting from the partition function representation (first equality of
Eq.~\ref{eq:B2_integral} in the main text),
\[
\btwo = \Vol-\frac{\QAB}{\QA\QB} \; ,
\]
we denote the isolated single-chain Hamiltonians in units of
$k_{\rm B}T$ ($k_{\rm B}$ is Boltzmann constant and $T$ is
absolute temperature) for $A$ and $B$,
respectively, as $\HH{A}[\R^A]$ and $\HH{B}[\R^B]$.
The corresponding conformational partition functions are then given by
\begin{subequations}
\begin{align}
\QQ_i = & \frac{1}{\Vol}\int \DD[\R^i] e^{-\HH{i}[\R^i]} \; , \; i = A,B \\
\QAB = & \frac{1}{\Vol}\int \DD[\R^A] \DD[\R^B] 
e^{-\HH{A}[\R^A]-\HH{B}[\R^B]-\U^{AB}[\R^A,\R^B]} \; ,
\end{align}
\end{subequations}
where $1/V$ cancels the degeneracy due to
translational invariance. It follows that
\begin{equation}
\begin{aligned}
\frac{\QAB}{\QA\QB} = & V\frac
	{\int \DD[\R^A]\DD[\R^B] e^{-\HH{A}[\R^A]-\HH{B}[\R^B]-\U^{AB}[\R^A, \R^B] }}
	{\int \DD[\R^A] e^{-\HH{A}[\R^A]} \int \DD[\R^B] e^{-\HH{B}[\R^B]}} \\
= & V \int \DD[\R^A]\DD[\R^B]
	\frac{ e^{-\HH{A}[\R^A]} }{\int \DD[\R^A] e^{-\HH{A}[\R^A]}}
	\frac{ e^{-\HH{B}[\R^B]} }{\int \DD[\R^B] e^{-\HH{B}[\R^B]}}
	e^{-\U^{AB}[\R^A, \R^B] } \\
\equiv & V\int \DD[\R^A]\DD[\R^B]\PP{A}[\R^A]\PP{B}[\R^B] e^{-\U^{AB}[\R^A, \R^B] }
\; ,
\end{aligned}
	\label{eq:Q_to_P}
\end{equation}
where, as noted in the main text,  
$\U^{AB}$ is in units of $k_{\rm B}T$,
the single-chain probability density function
\begin{equation}
\PP{i}[\R^i] \equiv \frac{ e^{-\HH{i}[\R^i]} }{\int \DD[\R^i] e^{-\HH{i}[\R^i]}} \; , \; i = A,B
\; ,
\end{equation}
and hence $\int \DD[\R^i]\PP{i}[\R^i] = 1$.
Substituting Eq.~\ref{eq:Q_to_P} for $\QAB/(\QA\QB)$ results in the second 
equality in Eq.~\ref{eq:B2_integral} of the main text, viz.,
\[
\btwo = \Vol \int \DD[\R^A] \DD[\R^B]\PP{A}[\R^A]\PP{B}[\R^B]\left( 1- e^{-\U^{AB}[\R^{A}, \R^B]}  \right) 
\; .
\]
We now proceed to decouple translational invariance from the internal 
degrees of freedom of the chain molecules by the following change of 
coordinates:
\begin{equation}
\{ \R^i_1, \R^i_2, \dots, \R^i_{N_i} \} \to  
\{ \R^i_1, \rdiff^i_1, \rdiff^i_2, \dots,\rdiff^i_{N_i-1} \}, \; 
\rdiff^i_s \equiv \R^i_{s+1}-\R^i_{s}
\; ,
\end{equation}
which allows all intramolecular residue-residue distances of chain $i$ be 
expressed solely in terms of $\rdiff^i$s:
\begin{equation}
\R^{i}_{s} - \R^i_t = \sum_{\tau=t}^{s-1}\rdiff^i_\tau \quad (s>t)
\; .
\end{equation}
Since the potential energy of an isolated chain molecule in homogeneous space
should depend only on the relative positions of its residues irrespective 
of the location of the chain's center-of-mass, the single-chain Hamiltonian
for chain $i$ should be a function of $\rdiff^i$s and independent of 
the position of any one single residue, which we may choose, without
loss of generality, as the position $\R^i_1$ of the first residue.
With this consideration, the partition functions $\QA$, $\QB$ can be 
rewritten as
\begin{equation}
\QQ_i = \frac{1}{\Vol}\int d\R^i_1 \DD[\rdiff^i] e^{-\HH{i}[\rdiff^i]} 
= \int \DD[\rdiff^i] e^{-\HH{i}[\rdiff^i]} \; ,
\quad  i = A, B,
\end{equation}
where $\DD[\rdiff^i] \equiv \prod_{s=1}^{N_i-1}d \rdiff^i_s$ and because  
$\int d\R^i_1/\Vol = 1$. 
For distances between residues on different chains,
\begin{equation}
\R^{AB}_{st} \equiv \R^A_s - \R^B_t 
= \sum_{\tau=1}^s \rdiff^A_\tau - \sum_{\mu=1}^t \rdiff^B_\mu 
+ \R^{AB}_{11}
\; ,
\label{eq:R_AB}
\end{equation}
where $\R^{AB}_{11} \equiv \R^A_1 - \R^B_1$.
Thus, the intermolecular interaction $\U^{AB}$ is a function of 
$\R^{AB}_{11}$ and $\rdiff^A$, $\rdiff^B$ (shorthand for 
$\{\rdiff^A\}=\{\rdiff^A_1, \rdiff^A_2, \dots,\rdiff^i_{N_A-1} \}$, 
$\{\rdiff^B\}=\{\rdiff^B_1, \rdiff^B_2, \dots,\rdiff^i_{N_B-1} \}$).
The partition function of the $A$-$B$ complex may then be expressed as 
\begin{equation}
\begin{aligned}
\QAB = & \frac{1}{\Vol} \int d\R^A_1 d\R^B_1 \DD[\rdiff^A] \DD[\rdiff^B] 
	e^{-\HH{A}[\rdiff^A]-\HH{B}[\rdiff^B]-\U^{AB}[\rdiff^A, \rdiff^B, \R^{AB}_{11}]} \\
= & \int d\R^{AB}_{11} \DD[\rdiff^A] \DD[\rdiff^B] 
	e^{-\HH{A}[\rdiff^A]-\HH{B}[\rdiff^B]-\U^{AB}[\rdiff^A, \rdiff^B, \R^{AB}_{11}]}
\; ,
\end{aligned}
\end{equation}
where the second equality follows from the change of variable 
$\{\R^A_1, \R^B_1 \} \to \{ \R^{AB}_{11}, \R^B_1 \}$ (Jacobian equals unity)
and the fact that $\int d\R^B_1/\Vol = 1$.
In terms of $\{\rdiff^i\}$, the single-chain conformational probability density
functions are given by
\begin{equation}
\PP{i}[\rdiff^i] = \frac{e^{-\HH{i}[\rdiff^i]}}{\int \DD[\rdiff^i] e^{-\HH{i}[\rdiff^i]}} \; , \quad i = A,B
\; .
	\label{eq:Pi_rdiff}
\end{equation}
To arrive at a physically more intuitive (but mathematically equivalent)
formulation, we may replace the $\R^{AB}_{11}$ distance between the first 
residues of the two different chains as an integration variable by
the $\Rcm^{AB}$ distance between
the centers of mass of the two chains while leaving all
$\{\rdiff^i\}$ variables unchanged. 
Since the center-of-mass distance is defined as
\begin{equation}
\begin{aligned}
\Rcm^{AB} = &
	\frac{\sum_{s=1}^{\NA}M^A_s\R^A_s}{\sum_{s=1}^{\NA}M^A_s} 
	- \frac{\sum_{t=1}^{\NB}M^B_t\R^B_t }{\sum_{t=1}^{\NB}M^B_t} \\
= & \R^{AB}_{11}
	+ \frac{\sum_{s=1}^{\NA}M^A_s\sum_{\tau=1}^{s-1}\rdiff^A_\tau}{\sum_{s=1}^{\NA}M^A_s} 
	- \frac{\sum_{t=1}^{\NB}M^B_t\sum_{\mu=1}^{t-1}\rdiff^B_\tau}{\sum_{t=1}^{\NB}M^B_t}
\; ,
\end{aligned}
	\label{eq:R11_Rcm}
\end{equation}
where $M^i_s$ is the mass of the $s$th residue in chain $i$, 
$|\partial \Rcm^{AB}/\partial \R^{AB}_{11}|=1$, and because
$\partial \rdiff^i_s/\partial \R^{AB}_{11}=0$ for $i=A,B$ and 
$s=1,2,\dots,N_i-1$, the Jacobian of this coordinate transformation
is unity. Hence, by integrating variable shift 
$d\R^{AB}_{11} \to d\Rcm^{AB}$, one obtains
\begin{equation}
\begin{aligned}
\frac{\QAB}{\QA\QB} = & \int d\Rcm^{AB}\DD[\rdiff^A]\PP{A}[\rdiff^A] \DD[\rdiff^B]\PP{B}[\rdiff^B] 
	e^{-\U^{AB}[\rdiff^A,\rdiff^B, \Rcm^{AB}]}	\\
	\equiv & \int d\Rcm^{AB} \Avg{  e^{-\U^{AB}[\Rcm^{AB}; \rdiff^A,\rdiff^B ]}  }_{A,B}
\; ,
\end{aligned}
\end{equation}
which leads immediately to the center-of-mass representation 
\[
\btwo = \int d\Rcm^{AB}\; \Avg{ 1- e^{-\beta \U^{AB}[\Rcm^{AB}; \R^A, \R^B]}  }_{A,B}
\]
given by Eq.~\ref{eq:btwo_Rcm} of the main text with the $\beta=1/k_{\rm B}T$
factor explicitly included.

\section{Derivation for $\btwo$ in terms of Mayer $f$-functions}

We now substitute the cluster expansion in Eq.~\ref{eq:expUAB-f}
of the main text,
\begin{equation}
e^{-\U^{AB}}-1 \approx \sum_{s=1}^{\NA} \sum_{t=1}^{\NB} f_{st} 
+ \sum_{s\geq t=1}^{\NA} \sum_{l\geq m=1}^{\NB} f_{sl} f_{tm} 
- \sum_{s=1}^{\NA}\sum_{t=1}^{\NB}f_{st}^2
\; 
	\label{eq:expUAB-f}
\end{equation}
(where $s\geq t, l\geq m$ in the second term on the right hand side
means that every term being summed is distinct),
into the $\btwo$ formula in Eq.~\ref{eq:B2_integral}
of the main text,
\[
\btwo = -V \int \DD[\R^A] \DD[\R^B]\PP{A}[\R^A]\PP{B}[\R^B]\left( e^{-\U^{AB}[\R^A,\R^B]} -1 \right)
\; ,
\]
to perform the $ \DD[\R^A] \DD[\R^B]$ integration for each of 
the three summation terms in Eq.~\ref{eq:expUAB-f}.
To do so, it is useful to first make the
$\{\R^i\} \to \{\rdiff^i\}\cup\{ \R^i_1 \}$ 
change of variables, then substitute the $\PP{i}[\rdiff^i]$ in 
Eq.~\ref{eq:Pi_rdiff} for $\PP{i}[\R^i]$ to rewrite 
Eq.~\ref{eq:B2_integral} of the main text as
\begin{equation}
\btwo = -\int d\R^{AB}_{11} \DD[\rdiff^A] \DD[\rdiff^B] \PP{A}[\rdiff^A]\PP{B}[\rdiff^B]
		\left( e^{-\U^{AB}[\rdiff^A,\rdiff^B, \R^{AB}_{11}]} -1 \right)
\; ,
	\label{eq:btwo_rdiff}
\end{equation}
where $\U^{AB}[\R^A,\R^B]\rightarrow\U^{AB}[\rdiff^A, \rdiff^B, \R^{AB}_{11}]$
by virtue of Eq.~\ref{eq:R_AB} because
$\U^{AB}[\R^A,\R^B]$ takes the form of $\U^{AB}[\{ \R^{AB}_{st}\}]$
and thus $f_{st}=f_{st}(\R^{AB}_{st})$.
Substituting Eq.~\ref{eq:expUAB-f} into Eq.~\ref{eq:btwo_rdiff},
\begin{equation}
\begin{aligned}
\btwo \approx & -\sum_{s=1}^{\NA} \sum_{t=1}^{\NB} 
	\int d\R^{AB}_{11} \DD[\rdiff^A] \DD[\rdiff^B] \PP{A}[\rdiff^A]\PP{B}[\rdiff^B] f_{st}(\R^{AB}_{st})  \\
& + \sum_{s=1}^{\NA}\sum_{t=1}^{\NB} 
	\int d\R^{AB}_{11} \DD[\rdiff^A] \DD[\rdiff^B] \PP{A}[\rdiff^A]\PP{B}[\rdiff^B] f_{st}^2(\R^{AB}_{st}) \\
& - \sum_{s\geq t=1}^{\NA} \sum_{l\geq m=1}^{\NB} 
	\int d\R^{AB}_{11} \DD[\rdiff^A] \DD[\rdiff^B] \PP{A}[\rdiff^A]\PP{B}[\rdiff^B]  
		f_{sl} (\R^{AB}_{sl}) f_{tm} (\R^{AB}_{tm}) \\
\equiv & \Bo + \Bto + \Btt.
\label{eq:B_s}
\end{aligned}
\end{equation}
Using the inverse of the Fourier-transformed matrix of Mayer $f$-functions
$\fk{\kk}_{st}$ defined in Eq.~\ref{eq:fst_k} of the main text,
\[
f_{st}(\rr) =  \int \frac{d^3k}{(2\pi)^3} \fk{\kk}_{st} e^{i\kk\cdot\rr } 
\; ,
\]
$\Bo$, $\Bto$, and $\Btt$ are evaluated.
First, a term in the summation over $s,t$ for $\Bo$ is equal to
\begin{equation}
\begin{aligned}
& -\int d\R^{AB}_{11} \DD[\rdiff^A] \DD[\rdiff^B] \PP{A}[\rdiff^A]\PP{B}[\rdiff^B] f_{st}(\R^{AB}_{st})  \\
= & -\int d\R^{AB}_{st} \DD[\rdiff^A] \DD[\rdiff^B] \PP{A}[\rdiff^A]\PP{B}[\rdiff^B] 
	\int \frac{d^3k}{(2\pi)^3} \fk{\kk}_{st}  e^{i\kk\cdot\R^{AB}_{st}}  \\
= & -\int \frac{d^3k}{(2\pi)^3} \fk{\kk}_{st} \int d\R^{AB}_{st} e^{i\kk\cdot\R^{AB}_{st}} \\
= & -\int \frac{d^3k}{(2\pi)^3} \fk{\kk}_{st} (2\pi)^3 \delta^3(\kk) \\
= & -\fk{\zero}_{st}
\end{aligned}
\end{equation}
because the $d\R^{AB}_{11} \to d\R^{AB}_{st}$ change in integration variable
for the interchain distance can be applied without affecting the 
integrations over $\PP{i}[\rdiff^i]$.
It follows from Eq.~\ref{eq:B_s} that 
\begin{equation}
\Bo\equiv = -\sum_{s=1}^{\NA} \sum_{t=1}^{\NB}  \fk{\zero}_{st}
\; .
	\label{eq:B2_f1}
\end{equation}
Second, every corresponding term for $\Bto$ is integrated 
by the same change of variable:
\begin{equation}
\begin{aligned}
& \int d\R^{AB}_{11} \DD[\rdiff^A] \DD[\rdiff^B] \PP{A}[\rdiff^A]\PP{B}[\rdiff^B] f_{st}^2(\R^{AB}_{st})  \\
= & \int d\R^{AB}_{st} \DD[\rdiff^A] \DD[\rdiff^B] \PP{A}[\rdiff^A]\PP{B}[\rdiff^B] 
	\int \frac{d^3k}{(2\pi)^3} \fk{\kk}_{st} e^{i\kk\cdot\R^{AB}_{st}}  \!\!
	\int \frac{d^3k'}{(2\pi)^3} \fk{\kk'}_{st}  e^{i\kk'\cdot\R^{AB}_{st}}  \\
= & \int \frac{d^3k}{(2\pi)^3}\frac{d^3k'}{(2\pi)^3}  \fk{\kk}_{st}   \fk{\kk'}_{st} 
	\int d\R^{AB}_{st} e^{i(\kk+\kk')\cdot\R^{AB}_{st}} \\
= & \int \frac{d^3k}{(2\pi)^3}\frac{d^3k'}{(2\pi)^3}  \fk{\kk}_{st}  \fk{\kk'}_{st}  (2\pi)^3\delta^3(\kk+\kk') \\
= & \int \frac{d^3k}{(2\pi)^3}  \fk{\kk}_{st}  \fk{-\kk}_{st}
\; .
\end{aligned}
\end{equation}
Therefore, by Eq.~\ref{eq:B_s},
\begin{equation}
\Bto = \sum_{s=1}^{\NA} \sum_{t=1}^{\NB} 
\int\frac{d^3k}{(2\pi)^3}
\fk{\kk}_{st}  \fk{-\kk}_{st} 
	= \int\frac{d^3k}{(2\pi)^3}
{\rm Tr}\left[ \hat{f}(\kk)\hat{f}^{\rm T}(-\kk)\right]
\; ,
	\label{eq:B2_f2-2}
\end{equation}
\nobreak
where the ``${\rm T}$" superscript on a matrix denotes transposing the given
matrix. 
Third, each of the terms in the summation for $\Btt$, involving
two residue pairs $(s^A, l^B)$ and $(t^A, m^B)$ satisfying the 
$s\geq t, l\geq m$ condition, can also be evaluated by a similar
change of integration variable. Because
\begin{equation}
\begin{aligned}
\R^{AB}_{sl} = & \R^{AB}_{11} + \sum_{\tau=1}^{s-1}\rdiff^A_\tau - \sum_{\mu=1}^{l-1}\rdiff^B_\mu 
= \R^{AB}_{11}  
	+  \left( \sum_{\tau=1}^{t-1} + \sum_{\tau=t}^{s-1} \right) \rdiff^A_\tau
	- \left( \sum_{\mu=1}^{m-1} + \sum_{\mu=m}^{l-1} \right) \rdiff^B_\mu \\
= & \R^{AB}_{tm} + \sum_{\tau=t}^{s-1} \rdiff^A_\tau - \sum_{\mu=m}^{l-1} \rdiff^B_\mu\; ,
\end{aligned}
\end{equation}
by making the $d\R^{AB}_{11} \to d\R^{AB}_{tm}$ change in integration 
variable, we obtain
\begin{equation}
\begin{aligned}
& \int d\R^{AB}_{11} \DD[\rdiff^A] \DD[\rdiff^B] \PP{A}[\rdiff^A]\PP{B}[\rdiff^B] 
	f_{sl}(\R^{AB}_{sl})f_{tm}(\R^{AB}_{tm})  \\
= & \int \!d\R^{AB}_{tm} \DD[\rdiff^A] \DD[\rdiff^B] \PP{A}[\rdiff^A]\PP{B}[\rdiff^B] 
	\int \!\frac{d^3k}{(2\pi)^3} \fk{\kk}_{sl} e^{i\kk\cdot\R^{AB}_{sl}}  \!\!\!
	\int\! \frac{d^3k'}{(2\pi)^3} \fk{\kk'}_{tm} e^{i\kk'\cdot\R^{AB}_{tm}}  \\
= & \int \frac{d^3k}{(2\pi)^3}\frac{d^3k'}{(2\pi)^3} \fk{\kk}_{sl}  \fk{\kk'}_{tm}
	\int d\R^{AB}_{tm} e^{i(\kk+\kk')\cdot\R^{AB}_{tm} } \\
 & \qquad\qquad \times
	\int \DD[\rdiff^A] \PP{A}[\rdiff^A] e^{i\kk\cdot\sum_{\tau=t}^{s-1} \rdiff^A_\tau}
	\int \DD[\rdiff^B] \PP{B}[\rdiff^B] e^{-i\kk\cdot\sum_{\mu=m}^{l-1} \rdiff^B_\mu} \\
= & \int \frac{d^3k}{(2\pi)^3}\frac{d^3k'}{(2\pi)^3} \fk{\kk}_{sl} \fk{\kk'}_{tm} (2\pi)^3\delta^3(\kk+\kk') 
	\Avg{ e^{i\kk\cdot(\R^{A}_s-\R^{A}_t)} }_A \Avg{ e^{-i\kk\cdot(\R^{B}_l-\R^{B}_m)} }_B \\
\equiv & \int \frac{d^3k}{(2\pi)^3} \fk{\kk}_{sl} \fk{-\kk}_{tm}\Pk{A}{\kk}_{st}\Pk{B}{-\kk}_{lm}
\; ,
\end{aligned}
\end{equation}
where 
\begin{equation}
\Pk{i}{\kk}_{st} = \int \DD[\rdiff^A] \PP{A}[\rdiff^A] e^{i\kk\cdot\sum_{\tau=t}^{s-1} \rdiff^i_\tau} 
= \int \DD[\R^i] \PP{i}[\R^i] e^{i\kk\cdot\left(\R^i_s-\R^i_t\right)} 
= \Pk{i{\rm T}}{-\kk}_{st} 
\; ,
\label{eq:Pk}
\end{equation}
$i=A,B$,
is the Fourier transformation of the intrachain residue-residue correlation 
function in Eq.~\ref{eq:Pk} of the main text.
$\Btt$ is then computed by rearranging the summation:
\begin{equation}
\begin{aligned}
\Btt = & -\sum_{s\geq t=1}^{\NA} \sum_{l\geq m=1}^{\NB} f_{sl} f_{tm} \\
= & -\frac{1}{2}\sum_{s, t=1}^{\NA} \sum_{l, m=1}^{\NB} f_{sl} f_{tm} 
	-\frac{1}{2} \sum_{s=1}^{\NA} \sum_{l=1}^{\NB} f_{sl}^2 \\
= &  -\frac{1}{2}\int \frac{d^3k}{(2\pi)^3}\Bigg\{{\rm Tr}
		\left[  \hat{f}(\kk)\hat{P}^B(-\kk)\hat{f}^{\rm T}(-\kk)\hat{P}^A(-\kk) \right]
	+ {\rm Tr}\left[ \hat{f}(\kk)\hat{f}^{\rm T}(-\kk)\right] \Bigg\}.
\end{aligned}
	\label{eq:B2_f2-1}
\end{equation}
The last equality in the above Eq.~\ref{eq:B2_f2-1} follows because we
have applied the last equality in Eq.~\ref{eq:Pk}, i.e.,
$\Pk{i}{\kk}_{st} = \Pk{i}{-\kk}_{ts}$, and Eq.~\ref{eq:B2_f2-2}.
Now, by combining Eqs.~\ref{eq:B2_f1}, \ref{eq:B2_f2-2}, and \ref{eq:B2_f2-1}, 
the cluster expansion expression for $\btwo$ up to $O(f^2)$ is given by
\begin{equation}
\begin{aligned}
\btwo \approx & \Bo + \Btt + \Bto \\
= & -\sum_{s=1}^{\NA} \sum_{t=1}^{\NB}  \fk{\zero}_{st} 
  -\frac{1}{2}\int \frac{d^3k}{(2\pi)^3}{\rm Tr}\left[  \hat{f}(\kk)\hat{P}^B(-\kk)\hat{f}^{\rm T}(-\kk)\hat{P}^A(-\kk)	
	 - \hat{f}(\kk)\hat{f}^{\rm T}(-\kk)\right]
\; ,
\end{aligned}
\end{equation}
which is reported in the main text as Eq.~\ref{eq:B2_general}.

$\null$

\section{Generating sequences with random charge patterns}

Random sequences for our charge pattern analysis are constructed as follows.
For each integer $i$ between 1 and 25, 40 random neutral sequences containing
$i$ positively charged residues (each carries $+1$ charge), $i$ negatively 
charged residues (each carries $-1$ charge), and $50 - 2i$ neutral residues 
(carry 0 charge) are generated by randomly permuting the array
$(+1,\dots,+1,0,\dots,0,-1,\dots,-1)$ with $+1$ and $-1$ each repeated $i$
times and 0 repeated $50 - 2i$ times to produce 1,000 random sequences.
1,000 random pairs of the sequences in this pool of 1,000 sequences are
then selected to investigate the correlation between $\jSCD$ and $\SCD$. 

$\null$

\section{Mathematical principles for negative SCD}

Here we present an efficient numerical method to address the possible sign(s)
of SCD values. Although a rigorous proof for sequences of all lengths
is still lacking, the analysis below, which covers sequences of lengths up to
1,001, should provide a practical guide as to whether {\em all} charge 
neutral sequences have a negative SCD, which is a remarkable observation that 
has so far been borne out empirically from sequences chosen to be 
studied in the literature.

\begin{figure}[ht!]
    \centering
    \includegraphics[width=\columnwidth]{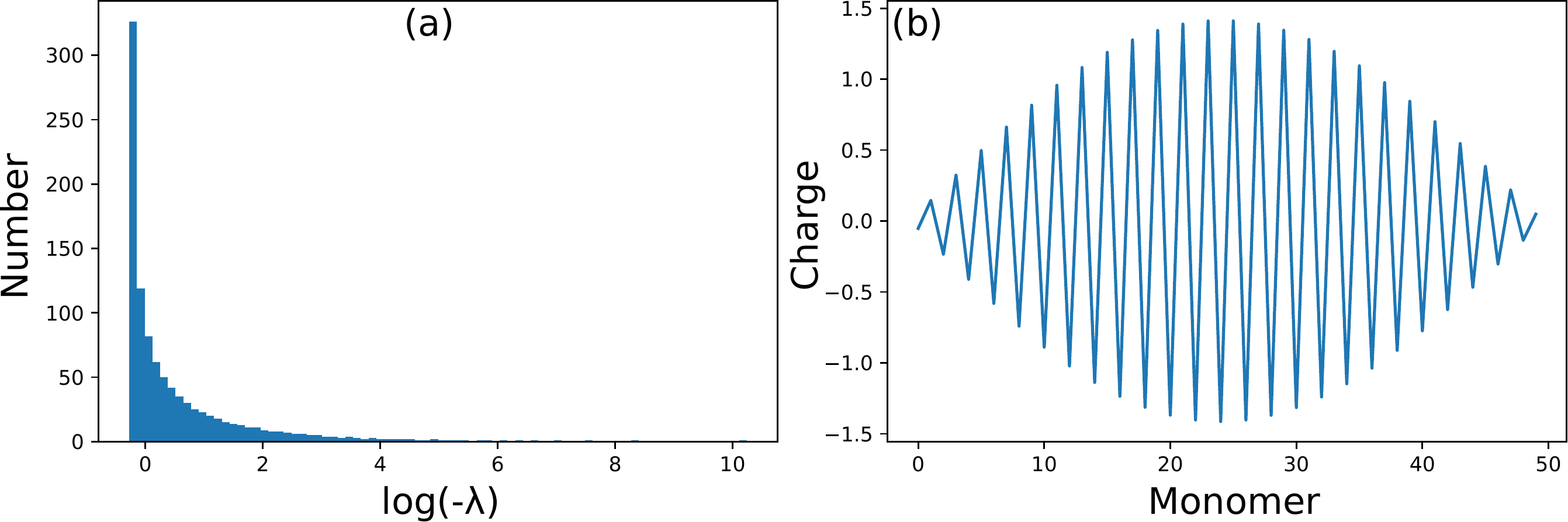}
    \caption{SCD value analysis. (a) The distribution of eigenvalues of
the matrix ${\hat B}_{1000}$ introduced in the text of this Supporting
Information for addressing the mathematical principles of negative $\SCD$
values for overall neutral sequences; all eigenvalues (denoted by $\lambda$)
shown are negative, 
demonstrating definitively that the $\SCD$ value of any overall charge neutral
sequence with equal or fewer than 1,001 residues is negative. The methodology
can readily be extended to test longer sequences insofar as
it is numerically feasible to determine the pertinent eigenvalues. 
(b) The charge distribution of a 50-residue, overall charge-neutral 
polyampholyte that produces the least-negative $\SCD$ value attained
numerically using gradient descent method.}
\label{fig:eigens}
\end{figure}

Consider a polymer of $N+1$ charges given by the column vector $\sigma$ $=$ 
$(\sigma_0,\sigma_1\dots,\sigma_N)$. By definition\cite{Sawle2015},
$\text{SCD}(\sigma)\equiv
\sum_{i=0}^N\sum_{j=i+1}^N\sigma_i\sigma_j\sqrt{|i-j|}$. 
If we define the
matrix ${\hat A}_{N+1}$ with elements $({\hat A}_{N+1})_{ij}=\sqrt{|i-j|}$, 
$\text{SCD}(\sigma)=\sigma^{\rm T} ({\hat A}_{N+1}/2) \sigma$.
If $\sigma$ is a charge pattern such that $\sum_{i=0}^N\sigma_i=0$, 
$\sigma_0=-\sum_{i=1}^N\sigma_i$. 
Now, defining $\bar \sigma=(\sigma_1,\sigma_2,\dots,\sigma_N)$ and the matrix
${\hat B}_N$ with elements
$({\hat B}_N)_{ij}=\sqrt{|i-j|}-\sqrt i -\sqrt j$, one can see that,
$\text{SCD}(\sigma)=\sigma^{\rm T} ({\hat A}_{N+1}/2) \sigma=
\bar \sigma^{\rm T} ({\hat B}_{N}/2) \bar \sigma$. Thus
the requirement that $\text{SCD}(\sigma)< 0$ for every $\sigma$ with
$\sum_{i=0}^N\sigma_i=0$ is equivalent to the requirement that 
$v^{\rm T} {\hat B}_{N} v< 0$ for any $N$-dimensional column vector $v$. 
It is a standard result of linear algebra
that, since ${\hat B}_N$ is self-adjoint, this is in turn equivalent to 
${\hat B}_N$ being a so-called ``negative matrix'', i.e., all of ${\hat B}_N$'s 
eigenvalues being negative. Notice
as well that for $M<N$, ${\hat B}_M$ is the top left $M\times M$ submatrix 
of ${\hat B}_N$, therefore, should ${\hat B}_N$ be negative, ${\hat B}_M$ 
would also be negative. For $N=1,000$, the
maximum (least-negative) calculated eigenvalue was about $-0.760$, confirming
that SCD is negative for neutral polymers at or under 1001 monomers. The
distribution of eigenvalues of ${\hat B}_{1000}$ is shown 
in Fig.~\ref{fig:eigens}a.
\\

{\bf\large Most charge-dispersed pattern (analyzed for $N=50$).}
Another quantity of interest is the smallest $-$SCD possible for a
neutral polyelectrolyte of some minimum nonzero charge (otherwise 
the totally neutral sequence in which every monomer carries 0 charge
would have the lowest $-$SCD at 0). In this regard, it is of interest to 
determine the lowest possible 
$\sigma^{\rm T} {\hat A}_{N} \sigma/{\sigma^{\rm T}\sigma}$ 
ratio for overall charge neutral $\sigma$ and the
charge pattern that produces it. The minimal value of this ratio 
produced by method of gradient descent is about $-0.761$, achieved by the
eigenvector with the charge distribution shown in Fig.~\ref{fig:eigens}b,
compared to about $-0.826$ for the strictly alternating 50-residue 
polyampholyte sv1.
\\

{\bf\large SCD values of non-neutral sequences.}
For a $N$-mer charge pattern $\sigma$ which is not necessarily overall
neutral, we can define its average charge 
$\langle\sigma\rangle\equiv\sum_{i=1}^N\sigma_i/N$ and represent
its sequence charge pattern by a column vector $p$ with components
$p_i=\sigma_i-\langle\sigma\rangle$. Thus we may write
$\sigma=p+\langle\sigma\rangle \mathbf{1}$ where $\mathbf{1}$ is the 
$N$-vector with a $1$ in every entry. Now we can express SCD as
\begin{equation}
\begin{aligned}
\text{SCD}(\sigma)&=\frac 1 2 \sigma^{\rm T} {\hat A}_{N} q\\
&=\frac 1 2 p^{\rm T} {\hat A}_{N} p+\langle\sigma\rangle p^{\rm T} 
{\hat A}_{N} \mathbf{1}+\frac 1 2 \langle\sigma\rangle^2 
\mathbf{1}^{\rm T} {\hat A}_{N} \mathbf{1}\\
&=\text{SCD}(p)+\langle\sigma\rangle\sum_{i=1}^Np_i( \sum_{j=1}^N\sqrt{|i-j|})+
\frac 1 2 \langle\sigma\rangle^2\sum_i^N\sum_j^N\sqrt{|i-j|}\\
&\approx \text{SCD}(p)+\frac 2 3\sigma\sum_{i=1}^Np_i\left[ i^{3/2}+(N-i)^{3/2}
\right] +\frac 4 {15} \langle\sigma\rangle^2N^{5/2}
\; ,
\end{aligned}
\end{equation}
where the last approximation 
follows by evaluating sums as integrals ($\sum_{z=1}^N\rightarrow\int_0^Ndz$).
$\text{SCD}(p)$ is negative as $p$ is overall charge neutral while 
$4 \langle\sigma\rangle^2N^{5/2}/15$ is, of course, positive and seemingly 
the primary contributor to increasing SCD for overall non-neutral sequences. 
As for the second (middle) term in the last expression, we note that
$[i^{3/2}+(N-i)^{3/2}]$ takes largest values when $i$ is low or high, 
i.e., when it represents monomers near the termini of the polymer sequence. 
It follows that $\langle\sigma\rangle\sum_{i=1}^Np_i[i^{3/2}+(N-i)^{3/2}]$ 
is positive if and only if the distribution of those monomers with charges 
of the same sign as that of the average charge $\langle\sigma\rangle$ is
biased in favor of being positioned at the two chain termini. 
In future studies, it would be interesting to explore
possible relationship between this finding and the recently discovered role 
of monomer type at chain termini in phase separation of model chains with
hydrophobic and hydrophilic monomers\cite{panag2020}
(labeled ``T'' and ``H'', respectively, and correspond essentially, in
that order, to 
the H and P monomers in the HP model~\cite{ChanDill1991,panag1992})
as well as the recently proposed ``SHD''
sequence hydropathy pattern measure for IDPs~\cite{Zheng2020}.

$\null$

\section{Explicit-chain simulation model and methods}

Coarse-grained molecular dynamics simulations are conducted for
six example pairs of $N=50$ sv sequences sharing a high-$\vert{\rm SCD}\vert$ 
sequence,
sv28, in common, that partners individually with six sv sequences 
spanning almost the entire range of charge patterns of
the 30 sv sequences. The pairs are sv28--sv1, sv28--sv10, sv28--sv15, 
sv28--sv20, sv28--sv24, and sv28--sv25.

We adopt the simulation model and method our group has recently applied
to study IDP phase separation~\cite{Das2018a,Das2018b}. 
Here, for simplicity, as in Ref.~\citen{Das2018b}, each residue (monomer) 
is represented by a van der Waals sphere of the same size and
mass. Each positively or negatively charged residue carries $+e$ 
or $-e$ charges, respectively, where $e$ is elementary electronic charge.
The potential energy function used for the study consists of
screened electrostatic, non-bonded Lennard-Jones (LJ) and bonded 
interactions. For any two residues $(i,s)$ and $(j,t)$---the $s$th 
residue of the $i$th chain and the $t$th residue of the $j$th chain---that 
carry charges $\ch{i}_s$ and $\ch{j}_t$, respectively, the 
residue-residue electrostatic interaction is given by
\begin{equation}
    U_{\rm el} = \frac{\ch{i}_s\ch{j}_t e^2}{4\pi\epsilon_0\epsilon_r 
r_{i,s;j,t}}\exp\left(- \kappaD r_{i,s;j,t}\right)
\; ,
    \label{eq:Uel_sim}
\end{equation}
where $\epsilon_0$ is 
vacuum permittivity, $\epsilon_r$ is relative permittivity,
and $r_{i,s;j,t}$ is the distance between residues $(i,s)$ and $(j,t)$.
We use $\kappaD=1/(3a)$
for the chain simulations in this work, where $a$ is a length unit
with roles that will be apparent below. If we take $a$ to correspond
roughly to the C$_\alpha$--C$_\alpha$ virtual bond length of $3.8$~\AA~
for proteins, $3a\approx 11$~\AA~ would be approximately equal to the 
Debye screening length for a physiologically relevant 150 mM aqueous 
solution of NaCl.
The non-bonded LJ interaction is constructed using the length scale $a$
as follows. Beginning with the standard LJ potential,
\begin{equation}
    U_{\rm LJ} = 4\varepsilon_{\rm LJ} 
    \left[\left(\frac{a}{r_{i,s;j,t}}\right)^{12}-
        \left(\frac{a}{r_{i,s;j,t}}\right)^6\right]
\; ,
\label{eq:LJ}
\end{equation}
where $\varepsilon_{\rm LJ}$ and $a$ are the depth and range of the 
potential, respectively,
we perform a cutoff and shift on Eq.~\ref{eq:LJ} to render the potential 
purely repulsive.
Since the main goal here is to compare explicit-chain simulation with
analytical theory, we use the non-bonded LJ part of the potential only for
excluded volume repulsion so that all attractive interactions in the
model arise from electrostatics as in the analytical theories considered
by this work. The final purely repulsive non-bonded LJ potential, 
$U_{\rm LJ}^{\rm cutoff}$ ($\geq 0$ for all $r_{i,s;j,t}$), 
that enters our simulation takes the Weeks-Chandler-Andersen 
form~\cite{weeks1971} 
\begin{equation}
U_{\rm LJ}^{\rm cutoff}= \left\{
\begin{array}{cc}
U_{\rm LJ} + \varepsilon_{\rm LJ} \; , &
\quad\quad {\rm for\ }
        r_{i,s;j,t} \leq 2^{1/6}a \\
    0 \; , & \quad\quad {\rm for\ } r_{i,s;j,t} > 2^{1/6}a
\end{array}
\right..
	\label{eq:ULJcut}
\end{equation}
As we have learned from Ref.~\citen{Das2018b}, the interaction among 
sv sequences can be strongly influenced by any background 
non-electrostatic interaction.
To make the energetics of our model system dominated by 
electrostatic interaction as in the analytical theories,
we set $\varepsilon_{\rm LJ} = \varepsilon/48$, where
$\varepsilon \equiv e^2/(4\pi\epsilon_0\epsilon_r a)$ is the electrostatic
energy at separation $a$, so that
short-range excluded-volume repulsion is significantly weaker 
than electrostatic interaction in the explicit-chain model. 
$\varepsilon$ and $a$ are used, respectively, as energy and length units
in our simulations.
As before, the bonded interaction between connected monomers is modeled using
a harmonic potential
\begin{equation}
    U_{\rm bond} = \frac{K_{\rm bond}}{2}\left( r_{i,s;i,s+1} - a \right)^2
\; ,
\end{equation}
with $K_{\rm bond} = 75,000\varepsilon/a^2$ 
as in Ref.~\citen{Silmore2017} and also our previous simulation of 
sv sequences~\cite{Das2018b}. The strength of this term is in line with the
TraPPE force field~\cite{Mundy1995,Martin1998,Nicolas2002,Pamies2003}.

All simulations are performed using the GPU version of HOOMD-blue simulation
package~\cite{Anderson2008,Glaser2015} at ten different 
temperatures (reported as reduced temperature 
$T^*\equiv k_{\rm B}T/\varepsilon = \lb/a$ 
for simulation results in this work) between
0.05$T^*$ and $0.5T^*$ with an interval of $0.05T^*$ using a timestep of
$0.001\tau_0$, where $\tau_0=\sqrt{ma^2/\varepsilon}$ is the reduced time 
defined by residue mass $m$.
For a given pair of sv sequences, simulation is initialized by
randomly placing the two chains in a large cubic box of dimension 
$100a\times100a\times100a$ then followed by $500\tau_0$ of energy minimization.
The electrostatic interactions among the residues are treated with the PPPM
method~\cite{LeBard2012} using a real-space cutoff distance of $15a$ and a
fixed Debye screening length of $3a$. After energy minimization, the system is
heated to its desired temperature in a time period of $2,500\tau_0$ 
using Langevin dynamics with a
weak friction coefficient of $0.1m/\tau_0$ (Ref~\citen{Silmore2017}). 
Motions of the residues are integrated using velocity-Verlet scheme 
with periodic boundary conditions. After the desired temperature is
achieved, a production run of 500,000$\tau_0$ is conducted and
trajectory snapshots are saved every $0.5\tau_0$ for subsequent analysis.

\section{Analysis of simulation data on binding}
\begin{figure}
    \centering
    \includegraphics[width=0.8\columnwidth]{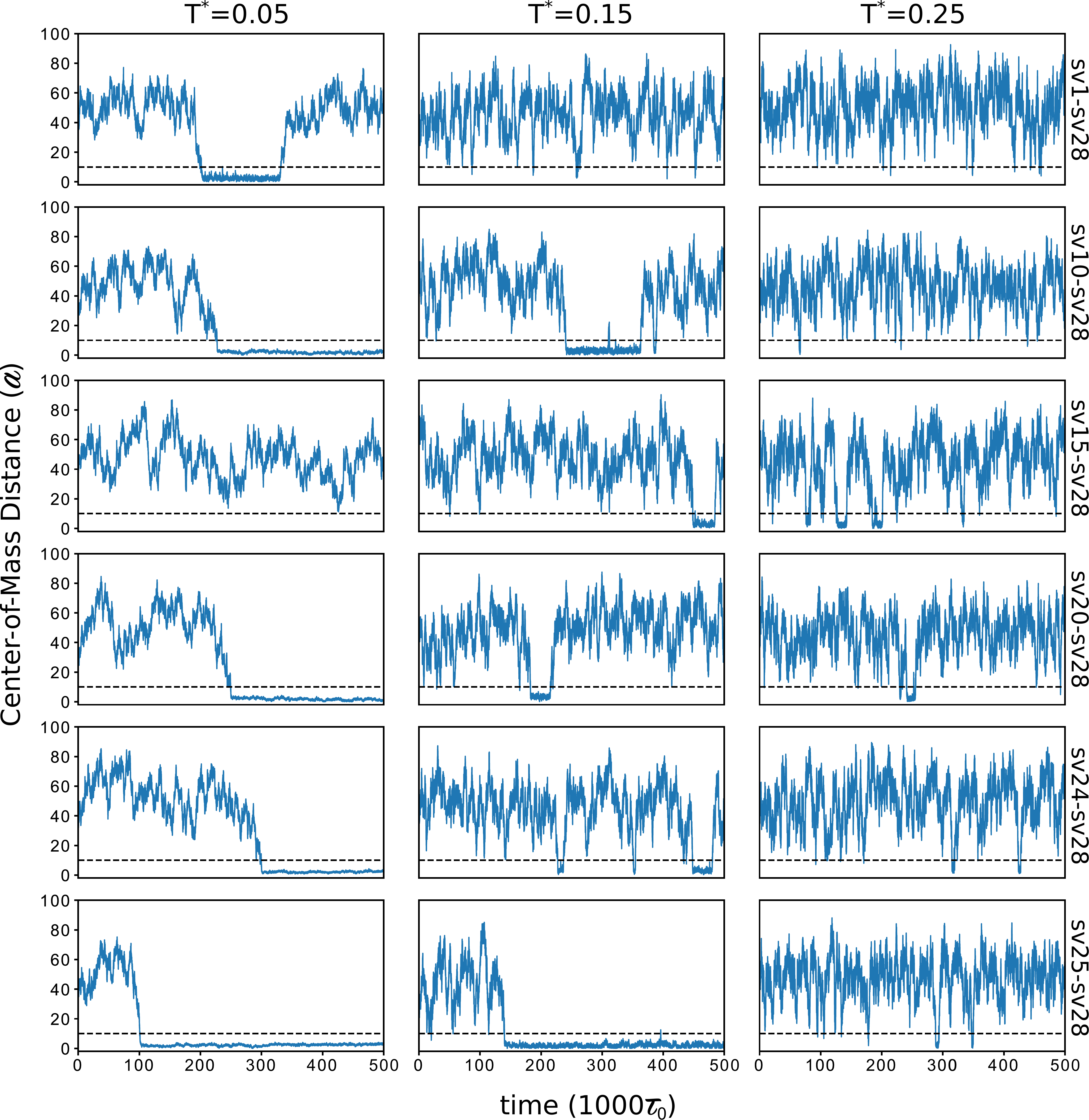}
\caption{Time dependence of the center-of-mass separation $|\Rcm^{AB}|$ between
the two sequences ($A$, $B$) in the explicit-chain simulations of sv 
sequence pairs at $T^*=0.05$, $0.15$, and $0.25$  
[$A=$ sv28, $B=$ (top to bottom) sv1, sv10, sv15, sv20, sv24, and sv25]. 
Dashed horizontal lines 
mark $|\Rcm^{AB}|=10a$, the cutoff adopted in the present work for 
identifying a ``bound state" of the two polyampholyte chains.  
None of the 18 center-of-mass distances plotted crosses the 
dashed lines more than five times, indicating potential limitations
in sampling under thermodynamic equilibrium conditions.}
\label{fig:poor_trajectory}
\end{figure}

For each simulation conducted for a given sv sequence pair, the simulated 
trajectory is examined for the center-of-mass separation between the two 
chain sequences to ascertain whether the chains form a binary complex in
each and all snapshot collected.
In the course of our investigation, we found that for simulations conducted 
at relatively low temperatures, $T^* < 0.35$, there were only very limited 
jumps between an unbound state and what would be reasonably considered as
the bound state (Fig.~\ref{fig:poor_trajectory}), suggesting that the 
simulated system may not have sufficient sampling at such low temperatures.
We therefore focus on simulations conducted at $T^* \geq 0.35$. 

Accordingly, the binding probabilities $\theta$ of the six pairs of 
sv sequences at $T^*=0.35$, $0.4$, $0.45$, and $0.5$ are calculated by
the method described in the main text.  As described there, we subtract
a constant baseline collision probability, $\theta_0$, of two noninteracting 
monomer, where $\theta_0=[4\pi(10a)^3/3]/(100a)^3$, from the 
simulated bound-state ratio, and use
$\tilde\theta=\theta-\theta_0$ to quantify the binding probability
produced by interaction energies.

Combining the simulation results from $T^*=0.35$, $0.4$, $0.45$, and $0.5$, 
we estimate an enthalpic parameter $\Delta H$ and an entropic parameter 
$\Delta S$ for the binding for each of the six sv sequence pairs 
using the linear 
regression
\begin{equation}
\Delta H/T^{*} -\Delta S=\log(\theta^{-1}-1)
\; ,
	\label{eq:theta_regress}
\end{equation}
the results of which are reported in
Table~\ref{tb:sim-regress}. The fitted $T^*$-dependent $\theta$s 
are then used to compute the corresponding $\tilde\theta$ values
at the same $T^*$ for all sv sequence pairs to compare with the
theory-predicted $\KD$s in 
Fig.~\ref{fig:Sim}c
and
Fig.~\ref{fig:KD_sim_all_reg} 
of the main text.

\newpage

\begin{table}[ht!]
    \centering
    \begin{tabular}{P{10cm}||P{4.5cm}}
    \hline
    \begin{tabular}{@{}P{1.5cm}|@{}P{2cm}@{}P{2cm}@{}P{2.5cm}@{}P{2cm}}
    \begin{tabular}{@{}c@{}}[NaCl]\\(mM)\end{tabular} & Theory &
    \begin{tabular}{@{}c@{}} Theory \\ H1-CTR\end{tabular} &
    \begin{tabular}{@{}c@{}} Theory \\ Net Charge\end{tabular} &
    ITC
    \end{tabular}
    &
    \begin{tabular}{@{}P{1.5cm}|@{}P{3cm}}
    \begin{tabular}{@{}c@{}}[NaCl]\\(mM)\end{tabular} & smFRET
    \end{tabular} \\
    \hline\hline
    \begin{tabular}{@{}P{1.5cm}|@{}P{2cm}@{}P{2cm}@{}P{2.5cm}@{}P{2cm}}
    165 & 3.41 & 4.59 & 142 & 0.46$\pm$0.05 \\ [1.4ex]
    220 & 5.09 & 6.77 & 189 & 0.72$\pm$0.03 \\ [1.4ex]
    260 & 6.46 & 8.55 & 223 & 2.0$\pm$0.1 \\ [1.4ex]
    300 & 7.94 & 10.46 & 257 & 6.1$\pm$0.1 \\ [1.4ex]
    350 & 9.95 & 13.06 & 300 & 9.6$\pm$0.7 \\
    \end{tabular}
    &
    \begin{tabular}{@{}P{1.5cm}|@{}P{3cm}}
    160 & $(2.1^{+1.1}_{-0.8})\times 10^{-6}$ \\
    180 & $(3.7\!\pm\!0.5)\times 10^{-5}$\\
    205 & $(1.0\!\pm\!0.1)\times 10^{-3}$\\
    240 & $(2.5\!\pm\!0.3)\times 10^{-2}$ \\
    290 & 0.23$\pm$0.15 \\
    330 & 0.14$\pm$0.04\\
    340 & 0.4$\pm$0.18 \\
    \end{tabular}  \\
    \hline
    \end{tabular}
    \caption{Theoretical and experimental ITC\cite{Feng2018}
and smFRET\cite{Borgia2018} $\KD$s (in units of $\mu$M) 
of H1-ProT$\alpha$ fuzzy complexes at different NaCl concentrations ([NaCl] 
in mM).
Amino acid sequences (in one-letter code) used in the theoretical 
calculation are taken from those studied by experiments, as follows 
(residues in red are not in the wildtype, they include
those remaining after proteolytic cleavage of the HisTag).
\\
ProT$\alpha$ (the ``ProT$\alpha$ (without His-tag)'' sequence in
Ref.~\citen{Feng2018}):\\ 
{\color{red}
GSYM}SDAAVDTSSEITTKDLKEKKEVVEEAENGRDAPANGNAENEENGEQEAD\\
NEVDEEEEEGGEEEEEEEEGDGEEEDGDEDEEAESATGKRAAEDDEDDDVDT\\
KKQKTDEDD;
\\
H1 (from Ref.~\citen{Feng2018}):\\
{\color{red}
M}TENSTSAPAAKPKRAKASKKSTDHPKYSDMIVAAIQAEKNRAGSSRQSIQKYIK\\
SHYKVGENADSQIKLSIKRLVTTGVLKQTKGVGASGSFRLAKSDEPKKSVAFKK\\
TKKELKKVATPKKASKPKKAASKAPTKKPKATPVKKTKKELKKVATPKKAKK\\
PKTVKAKPVKASKPKKAKPVKPKAKSSAKRAGKKK{\color{red}HHHHHH};
\\
H1-CTR (H1-C-terminal region, from Ref.~\citen{Borgia2018}):\\ 
SVAFKKTKKEIKKVATPKKASKPKKAASKAPTKKPKATPVKKAKKKLAATP\\
KKAKKPKTVKAKPVKASKPKKAKPVKPKAKSSAKRAGKKK{\color{red}GGPR}.\\
In the theoretical calculation, aspartic acid, glutamic acid (D, E) residues 
are each assigned $-1$ charge; arginine, lysine
(R, K) residues are each assigned $+1$ charge; all other residue types
are considered neutral ($0$ charge).
The ``Theory'' results in the table are calculated using 
both terms for $\btwo$ in Eq.~\ref{eq:B2_F1F2} of the main text,
whereas ``Theory Net Charge'' results are calculated using only the
first term in the same equation.
Because Eq.~\ref{eq:B2_F1F2} relies on the Gaussian-chain
approximation which may not be adequate
for the N-terminal globular domain of H1,
in addition to the data presented in Fig.~\ref{fig:Kd_H1_ProTa} of
the main text, we compute also $\KD$s for the binding between
the fully disordered C-terminal region of H1 (termed H1-CTR) with
ProT$\alpha$ using both terms for $\btwo$ in Eq.~\ref{eq:B2_F1F2} of
the main text and the 95-residue sequence for H1-CTR listed above. The 
resulting $\KD$s listed under ``Theory H1-CTR'' in this table
are about 1.2--1.5 times higher than those of full-length H1.
This difference in ProT$\alpha$ binding between full-length and 
C-terminal H1 is likely attributable to the subtraction of the $+18$ charges 
in its N-terminal domain~\cite{Borgia2018}.
}
    \label{tb:Kd_H1_ProTa}
\end{table}


\begin{table}[ht!]
    \centering
    \begin{tabular}{P{16cm}}

    \begin{tabular}{@{}P{2.15cm}|@{}P{2.15cm}@{}P{2.15cm}@{}P{2.15cm}@{}P{2.15cm}|@{}P{1.75cm}@{}P{1.75cm}@{}P{1.75cm}}
    \hline
    Sequence & $\theta|_{T^{*}=0.35}$ & $\theta|_{T^{*}=0.40}$ & $\theta|_{T^{*}=0.45}$ & $\theta|_{T^{*}=0.50}$ & $\Delta H$ & $\Delta S$ & $r^2$
    \end{tabular}

    \begin{tabular}{@{}P{2.15cm}|@{}P{2.15cm}@{}P{2.15cm}@{}P{2.15cm}@{}P{2.15cm}|@{}P{1.75cm}@{}P{1.75cm}@{}P{1.75cm}} 
    \hline
    sv1$\ $ & 0.362 \% & 0.432 \% & 0.420 \% & 0.252 \% & $-0.295$ & $-6.32$ & 0.225\\ [1.4ex]
    sv10 & 0.736 \% & 0.743 \% & 0.591 \% & 0.351 \% & $-0.810$ & $-7.08$ & 0.720\\ [1.4ex]
    sv15 & 1.01 \% & 1.64 \% & 0.923 \% & 0.803 \% & $-0.383$ & $-5.46$ & 0.202\\ [1.4ex]
    sv20 & 0.812 \% & 1.34 \% & 0.381 \% & 0.700 \% & $-0.594$ & $-6.33$ & 0.178\\ [1.4ex]
    sv24 & 4.04 \% & 1.83 \% & 2.56 \% & 0.976 \% & $-1.39$ & $-7.17$ & 0.703\\ [1.4ex]
    sv25 & 3.00 \% & 0.590 \% & 0.912 \%  & 0.228 \% & $-2.59$ & $-11.1$ & 0.787\\
    \hline
    \end{tabular}
    \end{tabular}
    \caption{Simulated binding data and regression parameters; $r^2$ is square 
of Pearson correlation coefficient of the regression.}
    \label{tb:sim-regress}
\vskip 15cm
\end{table}




\newpage

\newpage

\newpage



\setcounter{page}{1}
\renewcommand{\thepage}{R\arabic{page}} 

\bibliography{2IDP_interact}

\end{document}